%% file: main.tex
\documentclass[10pt,conference]{IEEEtran}
\IEEEoverridecommandlockouts

\usepackage{cite}
\usepackage{amsmath,amssymb,amsfonts}
\usepackage{algorithmic}
\usepackage{graphicx}
\usepackage{textcomp}
\usepackage{xcolor}
\usepackage[font=small]{caption}
\usepackage{afterpage}
\usepackage{microtype}
\usepackage{setspace}
\usepackage{enumitem}
\usepackage{xspace}
\usepackage{tcolorbox}
\usepackage{balance}
\usepackage{tabularx,ragged2e}
\usepackage{parnotes}
\usepackage{booktabs}
\usepackage{placeins}
\usepackage{stfloats}

\usepackage{hyperref}
\usepackage[all]{hypcap}    

\hypersetup{
colorlinks=false,
linkbordercolor=white,
urlbordercolor=white,
pdfborder={0 0 0}
}

\begin{document}

\title{FlexNN: A Dataflow-aware Flexible Deep Learning Accelerator for Energy-Efficient Edge Devices}

\author{\IEEEauthorblockN{Arnab Raha, Deepak A. Mathaikutty, Soumendu K. Ghosh and Shamik Kundu*}
\IEEEauthorblockA{\textit{Advanced Architecture Research, NPU IP, CGAI (CCG), Intel Corporation,} Santa Clara, CA, USA\\
Email: \{{arnab.raha, deepak.a.mathaikutty, soumendu.ghosh, shamik.kundu}\}@intel.com}
\vspace{-0.2in}
\thanks{*Shamik Kundu contributed to this work during his internship in the Advanced Architecture Research Group during summers '22 and '23. He is currently a PhD student at UT Dallas. (Email: \textit{shamik.kundu@utdallas.edu})}
}


\newcommand{\ie}{{\emph{i.e.,}}\xspace}
\newcommand{\viz}{{\emph{viz.,}}\xspace}
\newcommand{\eg}{{\emph{e.g.,}}\xspace}
\newcommand{\etc}{{\emph{etc.,}}\xspace}

\newcommand{\dnn}{{\textsc{FlexNN}}\xspace}
\newcommand{\flextree}{{\textsc{FlexTree}}\xspace}
\newcommand{\flexdrain}{{\textsc{FlexDrain}}\xspace}

\newcommand{\fx}{\textsc{Fx}\xspace}
\newcommand{\fy}{\textsc{Fy}\xspace}
\newcommand{\ic}{\textsc{Ic}\xspace}
\newcommand{\oc}{\textsc{Oc}\xspace}
\newcommand{\of}{OF\xspace}
\newcommand{\im}{IF\xspace}
\newcommand{\fl}{FL\xspace}
\newcommand{\fsape}{FSAPE\xspace}
\newcommand{\fsad}{FSAD\xspace}
\newcommand{\psum}{pSum\xspace}

\newcommand{\resnet}{ResNet50\xspace}
\newcommand{\mobilenet}{MobileNetV2\xspace}
\newcommand{\gnet}{GoogLeNet\xspace}
\newcommand{\inception}{InceptionV3\xspace}
\renewcommand{\sectionautorefname}{Section\xspace}
\renewcommand{\figureautorefname}{Fig.\xspace}

\setlength{\abovecaptionskip}{2ex}
\setlength{\belowcaptionskip}{2ex}
\setlength{\floatsep}{3pt}
\setlength{\textfloatsep}{3pt}

\graphicspath{{./figs}}

\maketitle

\pagestyle{plain}

\input{0_abstract}
\input{1_introduction}
\input{2_motivation}
\input{3_design}

\input{4_exp_method}
\input{5_results}
\input{6_related_work}
\input{7_conclusion}

\section{Acknowledgements}
We would like to sincerely thank Gautham Chinya, Debebrata Mohapatra, Huichu Liu, Moongon Jung, Sang Kyun Kim, Guruguhanathan Venkataramanan, Raymond Sung, Hong Wang, and Cormac Brick for their contributions to this work.

\bibliographystyle{IEEEtran}
\bibliography{main}

\end{document}

%% file: 0_abstract.tex
\begin{abstract}
This paper introduces FlexNN, a \textbf{Flex}ible \textbf{N}eural \textbf{N}etwork accelerator, which adopts agile design principles to enable versatile dataflows, enhancing energy efficiency. Unlike conventional convolutional neural network accelerator architectures that adhere to fixed dataflows (such as input, weight, output, or row stationary) for transferring activations and weights between storage and compute units, our design revolutionizes by enabling adaptable dataflows of any type through software configurable descriptors. Considering that data movement costs considerably outweigh compute costs from an energy perspective, the flexibility in dataflow allows us to optimize the movement per layer for minimal data transfer and energy consumption, a capability unattainable in fixed dataflow architectures. To further enhance throughput and reduce energy consumption in the FlexNN architecture, we propose a novel sparsity-based acceleration logic that utilizes fine-grained sparsity in both the activation and weight tensors to bypass redundant computations, thus optimizing the convolution engine within the hardware accelerator. Extensive experimental results underscore a significant enhancement in the performance and energy efficiency of FlexNN relative to existing DNN accelerators.

\end{abstract}

\begin{IEEEkeywords}

Deep neural network accelerator, flexible data flow, sparsity acceleration, energy efficiency.
\vspace{-5pt}
\end{IEEEkeywords}

%% file: 1_introduction.tex
\section{Introduction}
\label{sec:intro}

The landscape of machine learning is experiencing an unprecedented surge, with a multitude of artificial intelligence (AI) networks proposed along with the development of numerous hardware platforms dedicated to accelerating Deep Neural Network (DNN) inference tasks. As the field progresses, the complexity of DNNs continues to grow, resulting in handling of large amounts of tensor data that exhibit diverse shapes and dimensions across different layers of existing networks. Moreover, with the continuous introduction of new networks, the dimensions of these tensor data are in constant flux. Consequently, there is a pressing need to engineer hardware accelerators with the flexibility to efficiently process network layers of varying dimensions \cite{raha2023efficient, raha2021design}.

Furthermore, the proliferation of edge devices, including wearables, smart cameras, smartphones, and surveillance platforms, underscores the importance of \textit{energy efficiency} in the design of DNN accelerators \cite{ghosh2023energy}. Given that tensor data processing involves traversing multiple levels of memory hierarchy, minimizing data transfer while maximizing data reuse and resource utilization emerges as critical imperatives to improve the energy efficiency of DNN accelerators \cite{raha2021special}.

However, prevailing accelerators for DNN execution, such as Eyeriss~\cite{eyeriss} and TPU~\cite{tpu}, typically adopt custom memory hierarchies and fixed dataflows. These architectures dictate the sequence in which the tensor data for the activations and weights are moved to processing units to execute the tensor operations for each layer of the network. Although effective, these approaches may not fully exploit the potential for energy efficiency optimization inherent in more flexible hardware designs. Therefore, there is a growing interest in exploring novel architectures and strategies that can better adapt to the evolving demands of DNN inference while simultaneously improving energy efficiency across a range of edge devices and AI applications.

The energy consumption for each layer in DNN inference is heavily influenced by the movement of data across the memory hierarchy and the level of reuse within the processing units. Previous studies have endeavored to characterize energy efficiency through analytical models while stressing the importance of enabling flexibility in scheduling tensors of varying dimensions~\cite{reusekwon}. This flexibility involves optimizing the ordering, blocking, and partitioning of tensors to maximize reuse from the innermost memory hierarchy, where the energy cost per unit of data moved is minimized. However, most existing DNNs, such as ResNet, YOLO, VGG, and GoogLeNet, comprise tens to hundreds of layers, each with different preferences for scheduling to achieve energy optimality. Fixed-schedule DNN accelerators can only offer optimal data reuse and resource utilization for a subset of DNN layers, thus limiting overall energy efficiency. Moreover, these accelerators exhibit strong network dependencies, which poses challenges in adapting to the rapidly evolving landscape of DNNs. Existing DNN accelerator designs from both industry and academia predominantly employ fixed schedules, such as input stationary (IS), weight stationary (WS), output stationary (OS), non-local reuse (NR), and row stationary (RS) \cite{chen2016eyeriss1, chen2016eyeriss}. The fixed dataflow characteristic of these accelerators originates from their tensor data distribution modules, which perform addressing to on-die storage, data transfer to processing engine arrays, and data storage to SRAM banks in a predetermined manner.
As a result, these accelerators lack the flexibility to implement different schedules (\textit{i.e.}, dataflows). Although software solutions on general-purpose CPUs and GPUs can reshape and load tensor data, fixed-function accelerators do not support flexibility. FPGAs, although offering flexibility, cannot alter the hardware configuration during execution from one layer to another.

In contrast to previous approaches, this paper proposes \textit{a dataflow-aware flexible DNN accelerator} that leverages schedule information from DNN layers to adapt tensor data shape and internal compute configuration per layer. This enables the compiler to configure the DNN accelerator optimally for handling tensor operations based on tensor dimensions. The key advantage of our proposed accelerator design lies in its ability to switch among multiple schedules based on layer characteristics, thereby minimizing memory accesses for a given tensor operation and resulting in significant energy savings at the accelerator level.

To further enhance performance and increase energy efficiency in the accelerator, we capitalize on the inherent sparsity in DNNs. Due to the nature of DNNs, weights associated with the network are often ``sparse," which means that they contain a significant number of zeros generated during the training phase \cite{parashar2017scnn, ghosh2024harvest}. These zero-valued weights do not contribute to the accumulation of partial sums during multiply-and-accumulate (MAC) operations. Additionally, highly sparse weights cause activations to become sparse in subsequent layers of the DNN after passing through non-linear activation functions like ReLU. Furthermore, network quantization (INT8/INT4) for edge device inference also results in a high number of zeros in both weights and activations. This fine-grained unstructured sparsity in weights and activations offers potential for improved energy efficiency and processing speed in two ways: (1) MAC computation can be gated or skipped, and (2) weights and activations can be compressed to reduce storage and data movement. The former reduces energy consumption, while the latter reduces both energy consumption and processing cycles. However, designing DNN accelerators to harness these benefits from sparsity is challenging due to irregular access patterns, workload imbalances, and under-utilization of MAC-based processing elements \cite{chen2019eyeriss}. 
Hence, in this paper, we develop a novel sparsity acceleration logic capable of skipping computation of zero-valued compressed data while simultaneously identifying non-zero elements in both activation and weight tensors. This will facilitate the implementation of an efficient convolution engine in the hardware accelerator at the edge, enabling efficient utilization of resources and enhancing overall performance and energy efficiency.

In this paper, we introduce \dnn, a \textbf{Flex}ible \textbf{N}eural \textbf{N}etwork accelerator, designed with agile principles to support versatile dataflows, thereby enhancing energy efficiency. Recognizing that data movement costs significantly outweigh compute costs in terms of energy consumption, the flexibility in dataflow enables us to optimize data transfer per layer, leading to minimal data movement and reduced energy consumption, an advantage not achievable in fixed dataflow architectures. Furthermore, to further boost throughput and reduce energy consumption within the \dnn architecture, we propose an innovative sparsity-based acceleration logic. This logic harnesses fine-grained sparsity in both activation and weight tensors to bypass redundant computations, effectively optimizing the convolution engine within the hardware accelerator. In summary,  this paper makes the following contributions.

\begin{itemize}
    \item This paper introduces a novel DNN accelerator, \dnn, designed to be sensitive to dataflow, offering flexibility by integrating DNN layer scheduling insights. By dynamically adjusting tensor data shape and internal compute configuration for each layer, the accelerator allows the compiler to optimize its performance in handling tensor operations, tailoring its configurations based on tensor dimensions for diverse neural network architectures.

    \item We introduce a novel sparsity acceleration logic that capitalizes on the unstructured fine-grained sparsity present in incoming activation and weights, thereby expediting inference execution within the DNN accelerator. Data are maintained in a zero-compressed format to mitigate storage and data movement expenses. Weights and activations are mapped while considering sparsity to enhance reuse, thereby enhancing overall performance.

    \item Extensive experimental evaluations conducted on six distinct DNNs that span both image classification and object detection tasks highlight the transformative impact of our accelerator. Specifically, our architecture showcases substantial improvements over fixed-schedule accelerators for ResNet101 and YOLOv2, demonstrating up to $77\%$ and $62\%$ energy reduction over Eyeriss and TPU, respectively. Furthermore, our accelerator achieves notable sparsity improvements for four additional DNNs, namely \resnet, \mobilenet, \gnet, \inception. Across these benchmarks, \dnn achieves a speedup of $1.8\times$\nobreakdash--$3.3\times$ over dense accelerators and $1.7\times$\nobreakdash--$2.0\times$ over semi-sparse accelerators with weight-sparsity support. Sparsity support also provides $1.7\times$\nobreakdash--$3.0
    \times$ and $1.6\times$\nobreakdash--$1.8\times$ improvement in energy efficiency compared to dense and weight-sparse accelerator. These results underscore the profound impact of our accelerator in enabling efficient execution of sparse and compact DNNs, significantly enhancing both speed and energy consumption metrics.
\end{itemize}

The remainder of the paper is organized as follows. \sectionautorefname \ref{sec:motivation} delineates the need for flexible dataflow and efficient two-sided sparsity acceleration logic. \sectionautorefname \ref{sec:design} describes the microarchitectural details of the proposed \dnn accelerator. \sectionautorefname \ref{sec:exp_method} presents the experimental setup followed by the results in \sectionautorefname \ref{sec:results}. The prior art in this domain is described in \sectionautorefname \ref{sec:related_work}. Finally, \sectionautorefname \ref{sec:conclusion} concludes the paper.

%% file: 2_motivation.tex
\section{Motivation}
\label{sec:motivation}

In this section, we delve into the fundamental motivations driving the design and development of our accelerator architecture, focusing on two key aspects: the paramount importance of \textbf{flexibility} and the critical need for efficient \textbf{sparsity acceleration}. By addressing these critical considerations, our accelerator aims to revolutionize the landscape of deep learning (DL) hardware, offering unparalleled versatility and performance across a wide range of DNNs and applications. We explore how these foundational principles drive innovation and shape the architectural decisions that lead to the design of \dnn.

\subsection{Importance of flexibility}
\label{sec:schedule}

\begin{figure}[t!]
	\centering
	\includegraphics[width=\columnwidth]{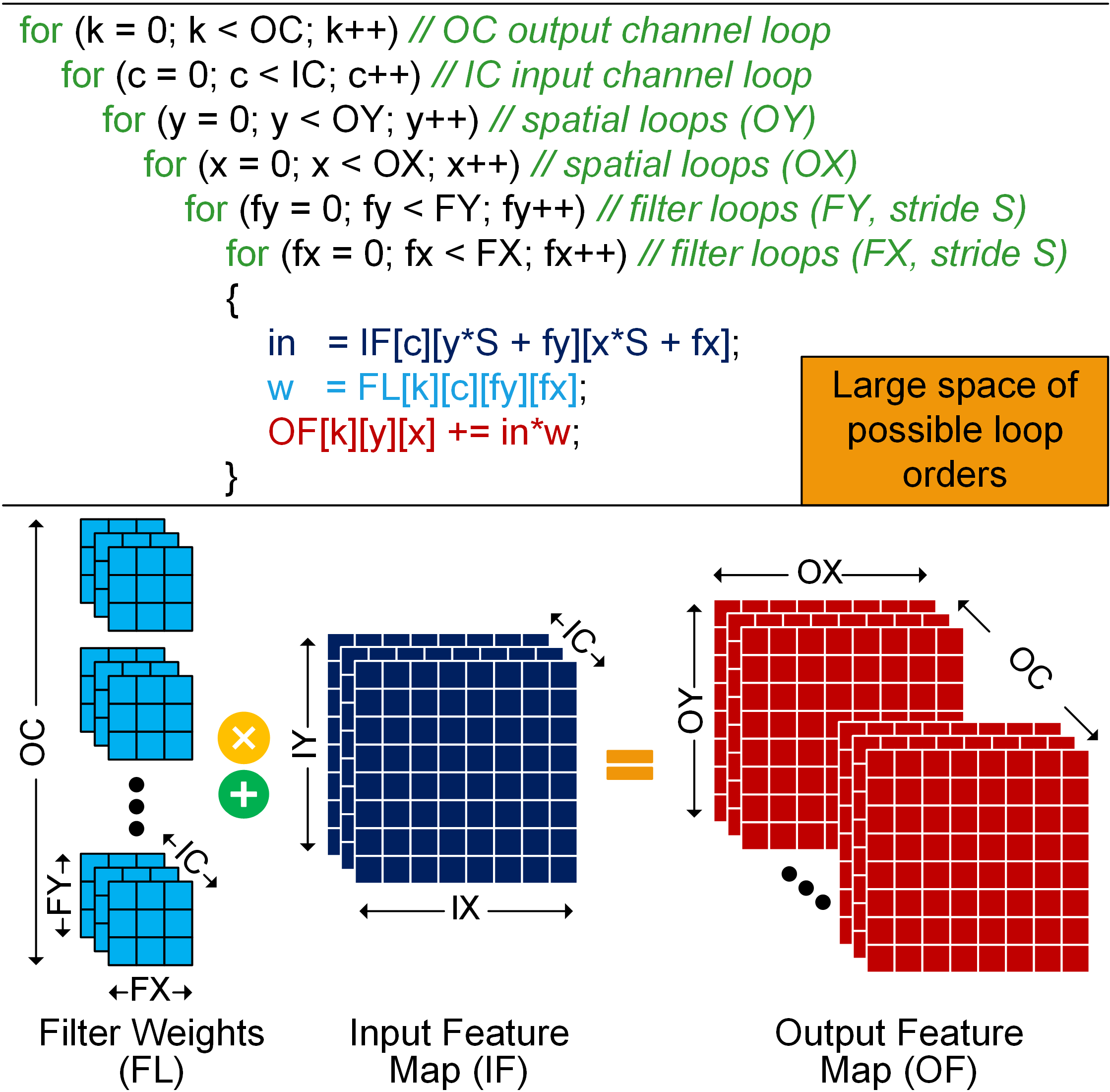}
        \caption{Illustration of multi-loop tensor processing during convolution operation in DNN.}
	\label{schedule}
\end{figure}

Numerous DNN accelerators utilize spatial architectures comprised of arrays of processing elements (PEs) alongside local storage, such as register files (RFs), for those PEs, and external storage, such as SRAM banks. In inference tasks, trained weights, or filters (FL), must be loaded into PE arrays from storage sources such as DRAMs and SRAM buffers. Input images, referred to as input activations or features (IFs), are also transferred to PE arrays, where MAC operations occur across multiple input channels (ICs) between activations and weights, generating output activations or features (OFs). Multiple sets of weight tensors (OCs) are commonly used against a specific set of activations to produce an output tensor volume. Finally, a non-linear function (\textit{e.g.}, ReLU) is applied to the output activations, which then become the input activations for the subsequent layer. Tensor processing involved in a convolution operation, as shown in \figureautorefname~\ref{schedule}, shows convolution layers comprising six nested loops. These layers generate an output tensor, OF map, from multiple kernel feature maps, FLs, operating on one or more input tensors, IF map. Each point in the output volume undergoes a MAC operation during the calculation. For instance, a 1$\times$1 convolution layer, such as the second convolution layer in ResNet50, illustrates IF map dimensions of IX = 56, IY = 56, IC = 64, and the filters dimensions of FX = 1, FY = 1, IC = 64, OC = 256. These dimensions convolve (with a batch size of 1) to produce an OF map with dimensions OX = 56, OY = 56, OC = 256, accompanied by appropriate padding values.

\begin{figure}[t]
	\centering
	\includegraphics[width=0.9\columnwidth]{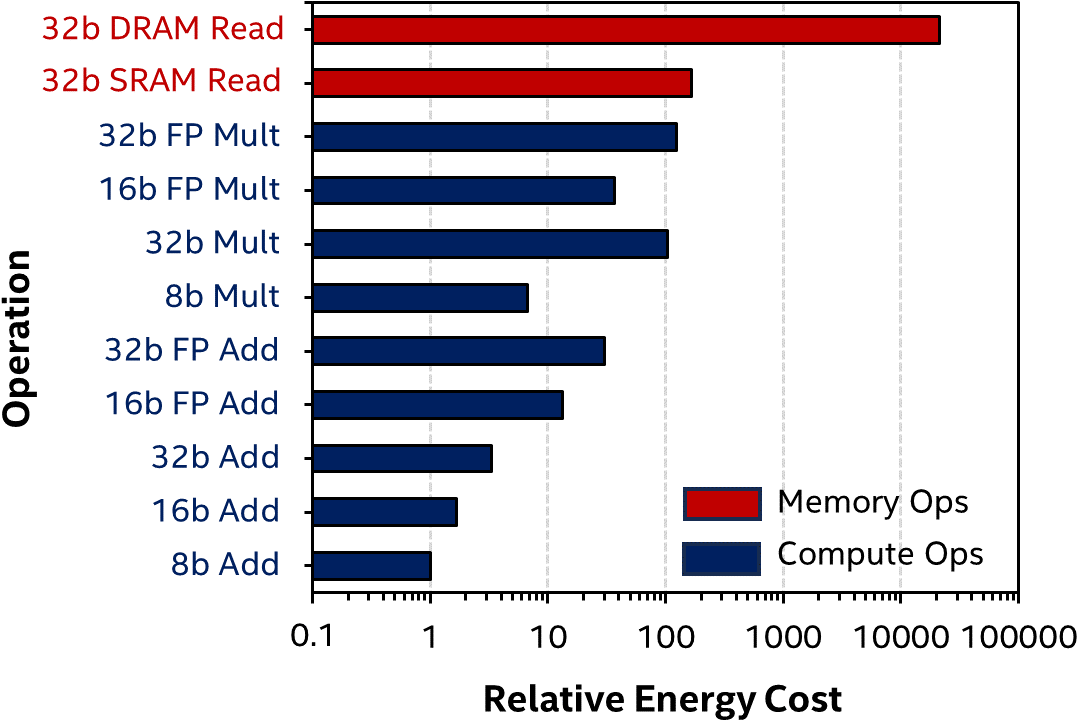}
	\caption{Relative energy costs of different compute and memory operations at various precisions in 45 nm technology\cite{cost}. Note that x-axis is in logarithmic scale.}
	\label{energy_cost}
\end{figure}

\begin{figure}[t]
	\centering
	\includegraphics[width=\columnwidth]{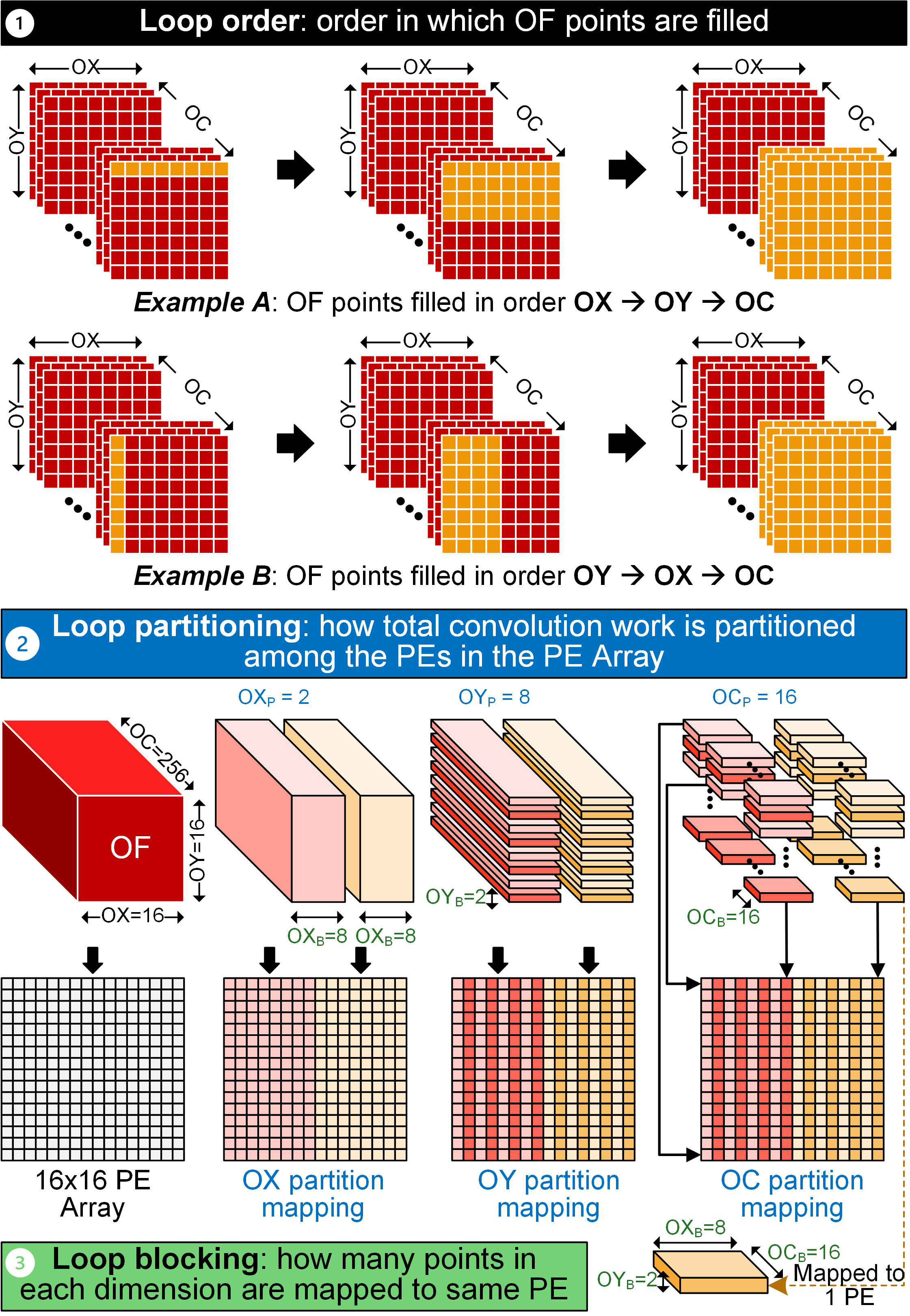}
	\caption{Illustration of (1) loop order, (2) loop partitioning and (3) loop blocking - referred to collectively as schedule - for optimizing data loading and distribution in the accelerator.}
	\label{schedule1}
\end{figure}

The dimensions of the input tensor undergo changes as they transition from one layer to another within a DNN and across various DNNs. Consequently, the development of flexible hardware accelerators becomes crucial to maintaining high utilization of compute units across network layers with arbitrary dimensions. Attempting to map various tensor dimensions to a fixed PE array with a consistent tensor mapping pattern can lead to decreased array utilization. To improve performance and energy efficiency, it is imperative to minimize data movement by maximizing data reuse from local memory and improving resource utilization. This optimization is particularly vital, as the cost of memory accesses often exceeds that of computing, as illustrated in \figureautorefname~\ref{energy_cost}. Numerous existing DNN accelerators, such as Eyeriss \cite{chen2019eyeriss}, TPU \cite{jouppi2017datacenter}, and SCNN \cite{parashar2017scnn}, implement novel memory hierarchies and fixed dataflows, influencing the movement of tensors for activations and weights within the processing units and the workload assigned to each PE. A fixed dataflow constrains the types of data movement across the memory hierarchy, limiting the degree of reuse within processing units. The movement of IFs, FLs, and partial sums (psums) , as well as the order of reuse, directly impact the energy consumption for each layer. In the literature\cite{reusekwon}, inference accelerators are classified into IS, WS, OS, and RS based on dataflow. The data reuse scheme is based on \textit{loop order}, \textit{loop blocking}, and \textit{loop partitioning} for tensor processing, collectively called a ``schedule'', as depicted in \figureautorefname~\ref{schedule1}. This schedule is described in relation to the dimensions of the tensors in a convolutional neural network. The loop order dictates the relative order of IX, IY (spatial), and IC dimensions for activations, and FX, FY, IC, OC dimensions for filters when loading these data into the accelerator. Loop partitioning dictates how the overall convolution operation is distributed among the PEs in the PE array, whereas loop blocking governs the allocation of multiple points in each dimension to the same PE.

All existing inference engines operate with fixed loop orders, blocking, and partitioning for convolution operations. Consequently, each accelerator can execute only one predetermined dataflow, where the data remain stationary in a single aspect. Various schedules require that IFs, FLs, and OFs/psums be mapped and accessed from local RF storage differently, depending on the type of schedule being computed. For example, in the IS scenario, a single point within the IF RF must undergo multiplication and accumulation against multiple points in the FL RF. The frequency of this repetition of operations varies on the basis of the schedule. Similarly, in the WS situation, a single point within the FL RF must be multiplied and accumulated against multiple points in the IF RF. Lastly, for OS schedules, the same psum in the OF RF must be retained and used to accumulate the results of multiplication between distinct IF and FL RF points over multiple cycles. Furthermore, when the size of the SRAM imposes limitations on the number of IC points that can be stored, incomplete OF points in the form of psums must be transferred to a higher-level memory hierarchy (\textit{e.g.,} DRAM) for subsequent retrieval into PE RFs to complete OF computation across all ICs.

Previous research aimed at characterizing the energy efficiency of DNN accelerators by constructing analytical models underscores the need to introduce flexibility in scheduling tensor operations of various dimensions to maximize reuse from the innermost memory hierarchy, where the energy cost per unit of data moved is minimized~\cite{reusekwon}. However, as mentioned in \sectionautorefname \ref{sec:intro}, fixed dataflows can only cater to optimal data reuse and resource utilization for a limited subset of DNN layers.  
To address this flexibility challenge, the proposed tensor data computing PE array offers a practical solution with minimal hardware overhead. Realizing a flexible dataflow accelerator requires a dataflow-aware tensor distribution unit capable of utilizing layer-specific optimal schedules and dataflow information to distribute data to the array. Moreover, the accelerator should inherently support flexible mapping and execution of these data within each PE.

\begin{center}
    \begin{tcolorbox}[colback=gray!10,
                      colframe=black,
                      width=8cm,
                      arc=1mm, auto outer arc,
                      boxrule=0.9pt,
                     ]
     \textbf{Motivation 1}:  \textit{It is important to develop a flexible dataflow in order to minimize data movement and maximize reuse in the PE array.}
    \end{tcolorbox}
\end{center}

\subsection{Importance of Sparsity Acceleration}

Sparse IFs are inherent in DNNs due to several factors. One primary cause is the prevalent use of ReLU as an activation function within many DNN architectures. The nature of ReLU to set negative values to zero contributes to sparsity, particularly intensifying in deeper layers, where it often exceeds $90\%$. In addition, the rise of auto-encoders, generative adversarial networks (GANs), and transformers further accentuates sparsity trends. These networks have decoder layers, employing zero-insertion techniques to up-sample input feature maps, resulting in more than $75\%$ zeros. Furthermore, extensive efforts have focused on inducing FL sparsity within DNNs. Various criteria, such as saliency, magnitude, and energy consumption, are used to determine which weights to prune. As a result, pruned networks exhibit weight sparsity levels of up to 90\% \cite{hoefler2021sparsity}.

The translation of the sparsity in weights and activations into enhanced energy efficiency and processing speed presents a significant opportunity. However, designing DNN accelerators capable of effectively harnessing these characteristics remains a formidable challenge. Computation gating emerges as a promising technique for converting sparsity in both IFs and FLs into energy savings. The implementation involves recognizing whether either the weight or activation is zero and clock-gating the datapath switching and memory accesses accordingly, achieving cost-effective solutions. To optimize throughput while conserving energy, skipping cycles of processing MACs with zero weights or activations becomes desirable. Yet, this necessitates intricate read logic to locate the next non-zero value without expending cycles on zeros. A natural solution entails maintaining FLs and IFs in a compressed format indicating the next non-zero location relative to the current one. However, compressed formats, often of variable length, pose challenges for parallel processing across PEs without compromising compression efficiency. Additionally, simultaneous recognition of sparsity in both weights and activations complicates matters, as efficiently `looking ahead' (\textit{e.g.}, skipping non-zero weights when the corresponding activation is zero) proves challenging with many compression formats. The irregularity introduced by such jumps precludes the use of pre-fetching to enhance throughput. Consequently, the control logic for processing compressed data becomes complex, adding overhead to the PEs. Addressing these complexities is vital to realize the full potential of sparsity in DNN accelerators.

As a result, hardware solutions in this domain have been limited. For example, Cnvlutin \cite{albericio2016cnvlutin} exclusively facilitates skipping cycles for activations without compressing weights, while Cambricon-X \cite{zhang2016cambricon} lacks the ability to maintain activations in compressed format. Given the intricacies involved in skipping cycles for both weights and activations, existing hardware designed for sparse processing tends to be tailored to specific layer types. For example, EIE \cite{han2016eie} is tailored for fully connected (FC) layers, while SCNN \cite{parashar2017scnn} is optimized for convolutional (CONV) layers. This specialization underscores the need for further innovation in developing versatile hardware architectures capable of efficiently handling sparsity across various layer types in diverse DNNs.

Introducing computation skipping for sparse data fundamentally alters the workload distribution across PEs, as the workload at each PE becomes contingent on sparsity levels. As the count of non-zero values fluctuates across diverse layers, data types, or even within specific regions within the same filter or feature map, it endangers an inherent imbalance in workload distribution across PEs \cite{chen2019eyeriss}. Consequently, the throughput of the entire DNN accelerator becomes constrained by the PE processing the highest number of non-zero MAC operations. This imbalance inevitably results in reduced PE utilization, thereby impeding the overall efficiency and performance of the DNN accelerator. Addressing this challenge requires innovative strategies to optimize workload distribution and improve PE utilization, thus maximizing the potential benefits of computation skipping for sparse data.

\begin{center}
    \begin{tcolorbox}[colback=gray!10,
                      colframe=black,
                      width=8cm,
                      arc=1mm, auto outer arc,
                      boxrule=0.9pt,
                     ]
     \textbf{Motivation 2}:  \textit{It is important to develop an acceleration logic that can leverage both unstructured IF and FL sparsity, in zero-compressed format.}
    \end{tcolorbox}
\end{center}

%% file: 3_design.tex
\section{Microarchitecture Design}
\label{sec:design}

\begin{figure}[t]
	\centering
	\includegraphics[width=1\columnwidth]{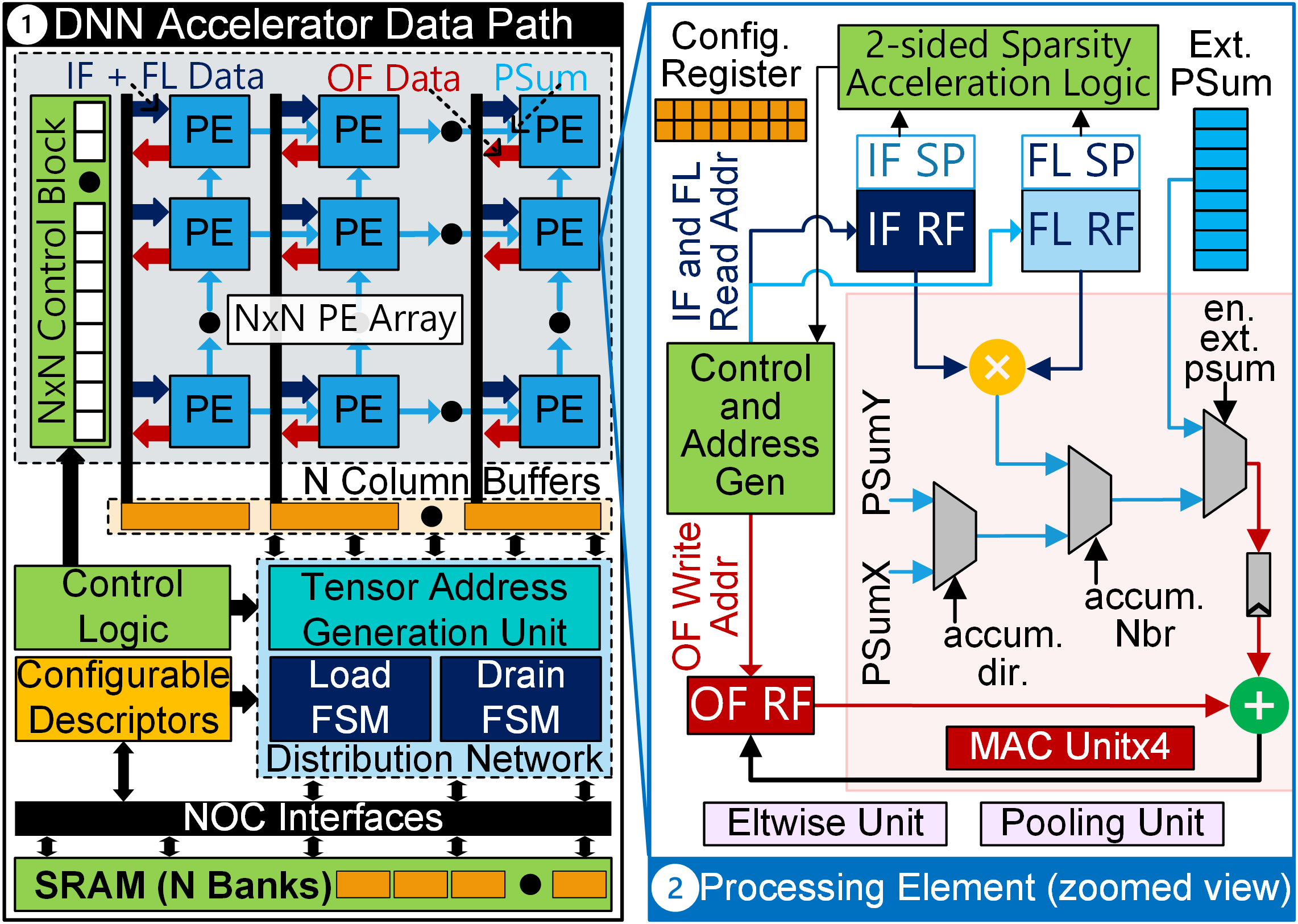}
	\caption{Top-level schematic of \dnn accelerator: (1) Interconnect pathways within PE array arranged in N columns, data distribution network, control block, and SRAM. (2) Architectural intricacies of each PE showing data storage (IF/FL/OF RF), sparsity storage (IF/FL SP), and MAC unit.}
	\label{toplevel}
\end{figure}

This section delineates the array of microarchitecture design decisions and techniques essential for implementing the proposed \dnn accelerator.

\subsection{Overview of \dnn accelerator}
\label{sec:dnnacc}

The high-level diagram of the DNN accelerator is shown in \figureautorefname~\ref{toplevel}, illustrating the various microarchitectural components that facilitate flexible and reconfigurable dataflow. Although the proposed DNN accelerator accommodates any schedule, it remains preferable to configure it according to the optimal schedule for individual layers of the neural network. The optimal schedule is obtained per layer, according to existing research \cite{yang2020interstellar, kwon2020maestro}. Leveraging the regularity of DNN computations facilitates efficient data loading into the accelerator and enables flexible convolution mapping based on the optimal schedule. In our design, we assume a three-level memory hierarchy. The first level comprises internal RFs within each PE. The second level consists of SRAM, similar to a small L1 cache, which stores input operands, output points, and partial sums. The third level encompasses DRAM memory, which is required because of its high capacity to store large amounts of filter weights and spilled OF points. For brevity, we omit DRAM-level implementation specifics and assume that the SRAM's capacity is sufficient to accommodate all input, intermediate activations, and filter weights. Each memory level gives us the opportunity to reuse data, thus enhancing energy efficiency. The flexible DNN accelerator, capable of supporting $INT8, U8, FP16$ and $BF16$ datatypes, comprises three principal components, elaborated in subsequent sections.

\subsection{Versatile Processing Element and Flexible Processing Element Array}
\label{sec:vpe}

The Versatile Processing Element (VPE) serves as the fundamental computational unit within the proposed \dnn accelerator, primarily tasked with performing MAC operations between IF and FL points \cite{mohapatra2020configurable, hsu2021multi, raha2021performance}. VPE also facilitates the accumulation of internal/external psums. In the context of flexible accelerator, the VPE optimizes the reuse of IF/FL/OF data and selects the most suitable compute template based on the optimal schedule for each layer. The Flexible Processing Element Array (FPA) comprises an N$\times$M array of VPEs, with the array dimension parameterized by design, typically through synthesis parameters. This array can be conceptualized as being arranged into M columns, each column consisting of N VPEs. To streamline the control logic, we deliberately adopted a square grid configuration (N$\times$N) for the FPA, using N = 16, simplifying the associated control mechanisms. 
\figureautorefname~\ref{toplevel}.1 illustrates the schematic of the DNN FPA, composed of an array of VPEs that serve as the most important computational unit within the accelerator. Although this figure shows an N$\times$N control block for simplicity, each PE column is instantiated with one control unit consisting of N control blocks, where each control block is dedicated to 1 PE. Each VPE, as demonstrated in \figureautorefname~\ref{toplevel}.2, features 4 sets of four 4R1W IF compressed data (CD) RFs to store input features ($IF_0 RF$ to $IF_3 RF$), 1R1W FL CD RF to store weights ($FL_0 RF$ to $FL_3 RF$), and 1R1W OF RF ($OF_0 RF$ to $OF_3 RF$) to store output values (OF/psum). In addition, each VPE consists of 4 sets of 1R1W RFs to store sparsity bitmaps (SP BMP), namely IF SP BMP RF and FL SP BMP RF. During a typical MAC operation, the input operands are fetched from the IF and FL RFs, based on the stored bitmaps, the sparsity acceleration logic (described later in \sectionautorefname~\ref{sec:csal}), and the addresses generated by \textit{control and address generation} (CAG) unit. The operation output is accumulated within the OF RF. Note that for stall-free high-performance execution, we introduced double buffers (active + shadow) in IF, FL, and OF RFs.

\begin{figure}[t]
	\centering
	\includegraphics[width=\columnwidth]{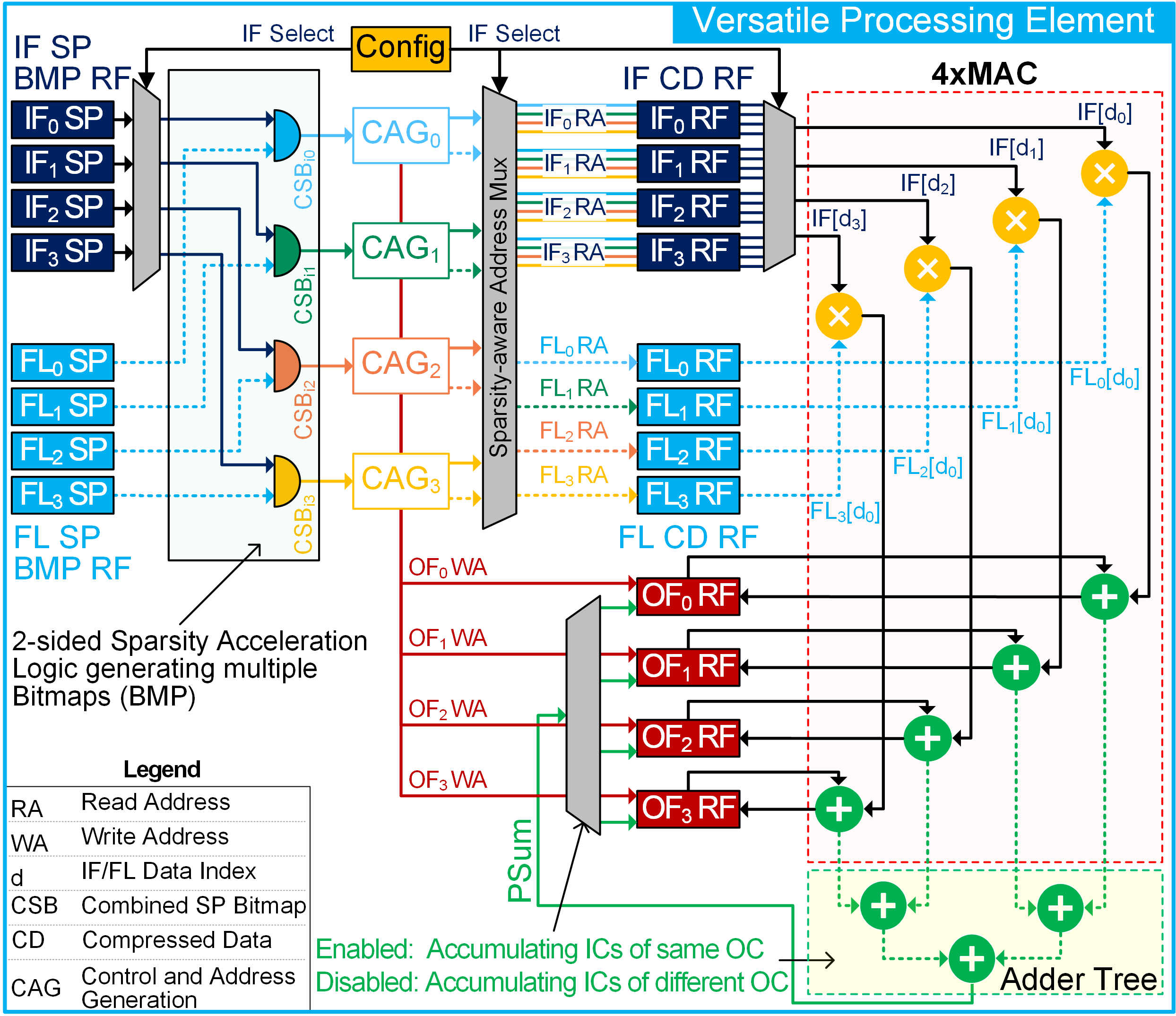}
	\caption{Microarchitecture of the \dnn VPE demonstrating inter-connectivity among control units, sparsity acceleration logic, sparsity bitmap registers (SP), data register files (RF), and MAC units, alongside support for two accumulation orders.}
	\label{flexarray}
\end{figure}

\figureautorefname\ref{flexarray} shows the microarchitecture details of VPE that execute computations on IF, FL, and OF/psum tensor data based on the optimal schedule of the current layer. VPE dynamically adjusts the loading and access patterns of the IF, FL, and OF/psum tensor data within the PE RFs to maximize reuse of the tensor data. The PE's microarchitecture is crafted to effectively utilize sparsity within both IF and FL. As illustrated in the figure, the PE comprises registers dedicated to storing sparsity data from incoming IF and FL streams, represented as bitmaps (IF SP BMP RF and FL SP BMP RF). These bitmaps are merged using a sparsity acceleration logic (further elaborated in \sectionautorefname\ref{sec:csal}), resulting in a combined sparsity bitmap (CSB). This unified bitmap serves as input to the CAG unit, facilitating the generation of participating non-zero IF and FL read addresses. N control blocks in each PE column updates the configuration descriptors inside individual PE at the onset of each convolution layer based on the optimal layer schedule, guiding data redirection during load, compute, and drain operations throughout the lifetime of the input tensor data. The PE finite-state machine (FSM) guides several internal counters and logic in the CAG unit to generate read and write control signals for IF, FL, and OF RFs, along with multiplexer control signals to route data from the RFs to the appropriate arithmetic units based on the template, \textit{viz.,} vector-vector (V$\times$V) or matrix-matrix (M$\times$M) or operation type, \textit{viz.,} MAC/Eltwise/Pooling and tensor dimension.

\begin{figure*}[t]
	\centering
	\includegraphics[width=\textwidth]{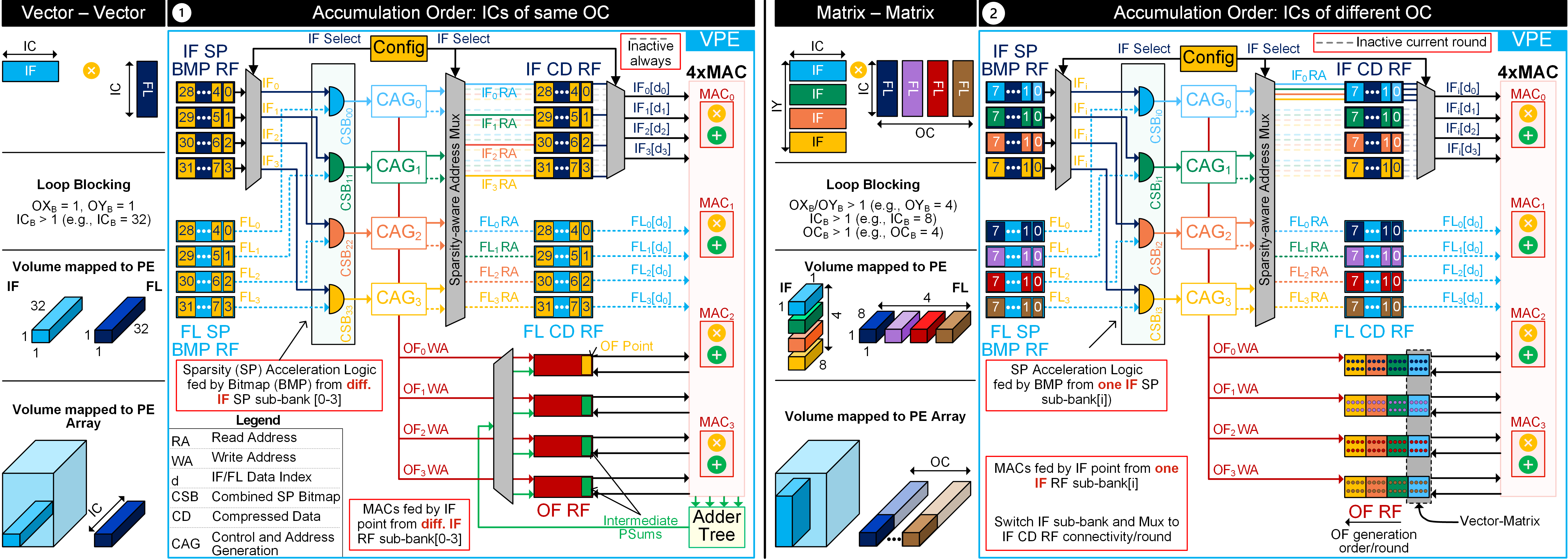}
    \caption{Versatile Processing Element (VPE) accommodating (1) V$\times$V and (2) M$\times$M templates. In V$\times$V, accumulation involves Input Channel (IC) of the same Output Channel (OC) of weights (FL), while M$\times$M entail accumulation of ICs from different OCs. M$\times$M involves different V$\times$M each round. This illustration presents sparsity-aware flexible dataflow inside the VPE along with loop blocking and partitioning. Weight filter dimensions assumed: FX=1 and FY=1.}
    \vspace{-15pt}
	\label{templates}
\end{figure*}

Assisted by the PE FSM, internal registers within the CAG unit track the total number of PE blocks (or OF/psum points) produced, aiding in addressing IF/FL/OF RFs. Additionally, counters like \textit{ifcount}, \textit{wcount}, and \textit{ofcount} manage the addresses/indexes for IF, FL, and OF RFs, increasing or clearing based on the number of input activations and weights required to calculate each OF point or psum block. The layer schedule determines the type and extent of IF/FL/OF RF data reuse, regulated by internal IF/FL/OF block counters controlling the loading of new IF/FL data and draining OF data each round, as per the layer's optimal schedule. These internal structures and associated control logic are crucial to supporting flexible schedules within the VPE. The critical role of VPE in facilitating flexible blocking within the DNN accelerator is realized by dividing the RF into multiple subbanks (X) and incorporating X MACs (\textit{e.g.}, X = 4) alongside multiplexers, allowing the implementation of V$\times$V, V$\times$M, and M$\times$M templates \cite{mohapatra2022runtime}, based on the optimal blocking factor of the layer ($IC_B, OC_B, OX_B, OY_B$), as shown in \figureautorefname~\ref{templates}.

For specific schedules, the convolution operation can be partitioned to split across multiple VPEs based on the number of ICs. Consequently, DNN computations that generate psums across different sets of ICs for a particular OF point should be mapped to a single column or row of VPEs. External psum accumulation enables accumulation of all ICs partitioned into multiple VPEs to generate the final OF point. The FPA facilitates the transmission of psums formed within the PE to its right or top neighbor, which is essential for the psum accumulation in FPA. To mitigate wire congestion and routing complexity, interconnections between PEs are restricted to their top and right neighbors. As shown in \figureautorefname\ref{toplevel}.2, three multiplexers are used with control signal \textit{accum\_dir} selecting the neighbor, \textit{accum\_Nbr} and \textit{en\_ext\_psum} selecting between external accumulation and internal MAC accumulation. Note that these multiplexers are not shown in \figureautorefname \ref{flexarray} and \ref{templates} for clarity. This architectural decision inherently influences how work is partitioned among different PEs, mainly in the IC dimension.

In certain optimal schedules, all ICs are not accumulated simultaneously. Instead, a portion of the IC set is initially loaded into the PE RFs, and the computed psum is extracted to the SRAM (or even DRAM) to be brought back into the PE RFs later when the remaining ICs are accumulated. External partial sum accumulation necessitates a 32-bit wide read and write direct bypass to and from the SRAMs. Sharing arithmetic units for MAC and Eltwise computation, along with multiplexer control logic routing appropriate tensor data into these units, reduces area overhead by enhancing hardware reuse efficiency within the PE. Residual networks such as ResNet require element-wise operations, such as the addition of OFs from two convolution layers. To support such operations while maximizing hardware resource reuse, OFs from two different layers are routed into the PE, using existing load and drain paths. The Eltwise field in the programmable descriptor signals an eltwise operation, bypassing the multiply operation within the PE and performing an eltwise addition of the two IF inputs.

\noindent
\textbf{Illustrative Example:} The \dnn PE demonstrates flexibility by executing V$\times$V and M$\times$M MAC operations, exploiting sparsity in both scenarios, as illustrated in \figureautorefname\ref{templates}.1 and 2 respectively. The PE adapts this flexibility in its operation based on the optimal schedule selected for each specific layer. Let us first delve into the V$\times$V operation scenario, where ICs within the same OC are accumulated. In this example with $IC_B=32$, 32 IFs and FLs are assigned to the PE for computation, corresponding to 32 distinct ICs but belonging to the same OC (represented by a single yellow color). Since these values exhibit sparsity, the sparsity bitmaps of IF and FL are stored in the respective registers IF SP BMP RF and FL SP BMP RF. The IF select signal retrieves the bitmaps from the first IF register ($IF_0$) and the first FL register ($FL_0$), transmitting them to the two-sided combined sparsity acceleration logic, which produces $CSB_{00}$. This logic identifies the non-zero activations and weight addresses through the CAG unit. These addresses guide the IF and FL CD RFs that store zero-compressed IF and FL values and provide precise values for MAC operations. Simultaneously, this process repeats for the other IFs ($IF_1$ to $IF_3$) and FL registers ($FL_1$ to $FL_3$), thus feeding the MACs with IF/FL points from different IF/FL RF subbanks and generating four psums concurrently, in each case, within the OF RFs. These psums aggregate to produce a single OF point upon completion of the computation. This cycle is iterated until all 32 ICs are processed. Subsequently, the next set of 32 ICs is loaded, and this process continues until all OF points for that OC are computed.

Now, let us dive into the second scenario of M$\times$M operations, focusing on the computation of ICs across different OCs within the PE. Here, IF SP BMP RFs and IF CD RFs receive bitmaps and input features that correspond to four distinct OCs, each represented by a different color. Similarly, the corresponding FL SP BMP RFs are loaded with four different FLs. During computation, in the initial round ($i$), only elements from the first IF SP BMP RF are provided as input to all CAGs along with four different FL bitmap values for the four OCs. Consequently, the CAG generates addresses solely corresponding to $IF_0$ following a two-sided sparsity acceleration logic ($CSB_{00}-CSB_{03}$). Subsequently, after obtaining the participating non-zero IFs and FLs from the compressed RFs, the MAC operation partially generates each of the four OC points in the first round, each denoted by a distinct color. Notably, in this case, MACs draw input exclusively from one IF RF subbank at a time, unlike the previous scenario in which they obtained IFs from all subbanks simultaneously. In the subsequent round, IF RF subbanks are switched using appropriate MUXing logic, consequently switching the compressed IF RF bank to acquire non-zero IF points for that specific round. Thus, contingent on the optimal layer schedule, the sparse PE efficiently conducts both V$\times$V and M$\times$M operations, capitalizing on both-sided sparsity in activations and weights.

\subsubsection*{Schedule-Aware Flexible Depth Adder Tree (\flextree)}
\label{sec:flextree}

Our \dnn accelerator's core features a tree-based architecture named \flextree, designed for psum accumulation across numerous PEs within the FPA to generate the final output point \cite{mohapatra2022schedule}. The distinguishing feature of \flextree is its ability to dynamically adapt the depth of the adder tree, allowing the compiler to create flexible schedules for network layers of varying dimensions. This hardware enhancement allows the compiler/scheduler to discover highly compute-efficient schedules. Before delving into the \flextree architecture, it is essential to understand the concept of Input Channel Partition ($IC_P$), similar to OF channel partition as illustrated earlier in \figureautorefname~\ref{schedule1}. $IC_P$ denotes how many ICs are assigned to a single PE in the FPA. Consequently, this also denotes the number of PEs that participate in the partial sum accumulation. Let us elucidate $IC_P$ using an example of 64 ICs. When $IC_P=1$, the computation involves only one PE, denoted PE1. All 64 ICs undergo pointwise multiplication and accumulation within PE1, producing the final output. When $IC_P=2$, 64 ICs are evenly divided between PE1 and PE2, each processing 32 ICs. PE1 accumulates psums from channels 0 to 31, while PE2 accumulates those from channels 32 to 63, forming the final output collectively. Similarly, for $IC_P=4$, the channels are distributed in PE1, PE2, PE3, and PE4, each PE handling 16 ICs. These psums of the respective sets of ICs are accumulated within each PE to generate the final output. Essentially, $IC_P\times IC_B=IC$. 

\begin{figure}[t]
	\centering
	\includegraphics[width=0.97\columnwidth]{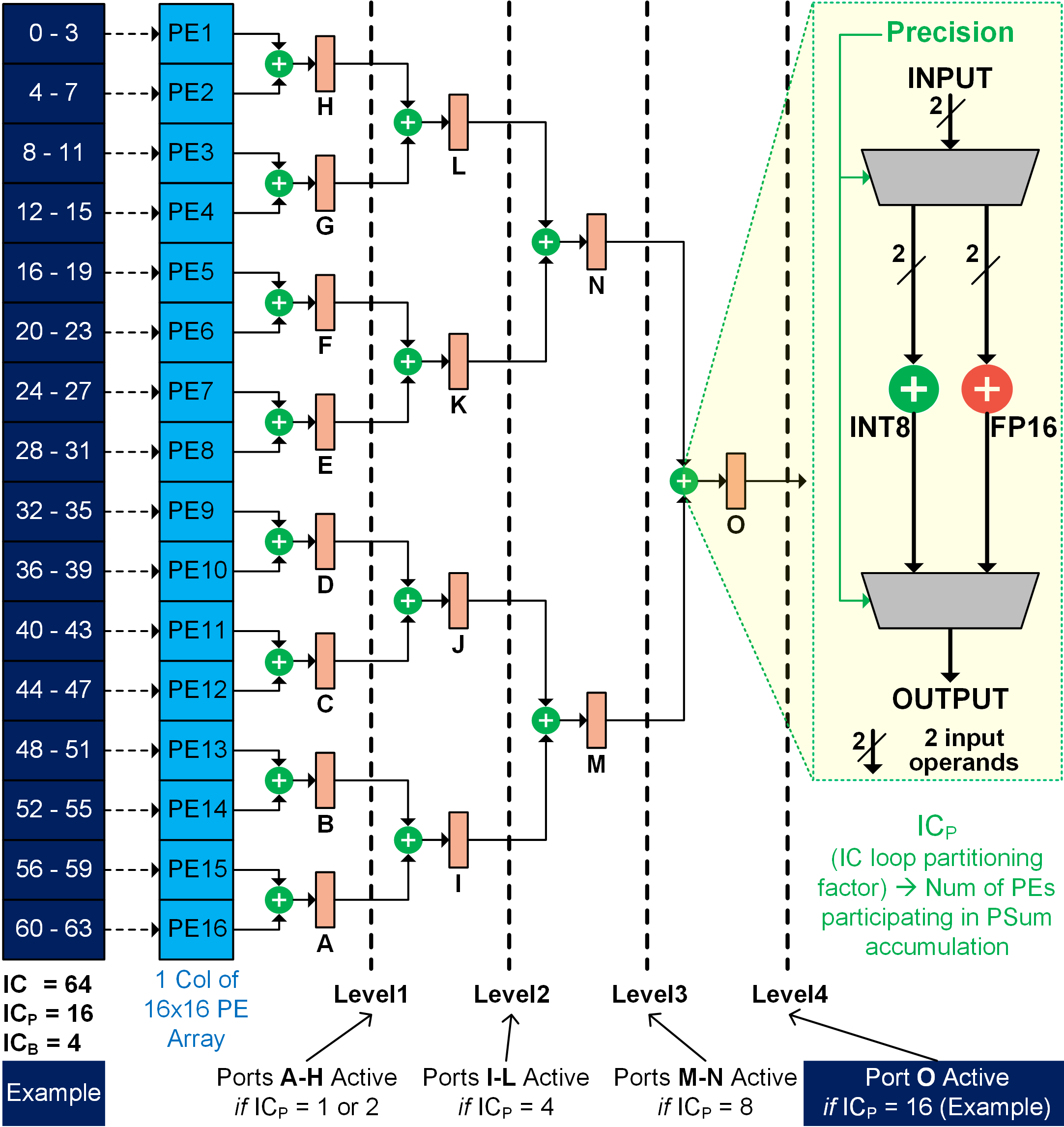}
	\caption{FlexTree architecture details with illustration using 64 input channels and 16 input channel partition factor.}
	\label{flextree}
\end{figure}

\figureautorefname~\ref{flextree} illustrates the \flextree architecture, which receives 16 inputs from the 16 PEs within a column of the PE array in the DNN accelerator. $IC_P$ supported by the adder tree network ranges from 1 to 16, inclusively. Even if $IC_P = 2$, the output of the computation must still pass through the adder tree network before producing the final OF output. This ensures a reduction in hardware overhead by simplifying hardware design and achieving uniformity across all $IC_P$ values. It is noteworthy that our \flextree architecture can accommodate $IC_P$ values that are not powers of 2 by entering zeros into the \flextree network of PEs that do not align with powers of 2. Each module marked with a `+' sign comprises both the INT8 adder and the FP16 adder to support convolution layers of different precision \cite{raha2022floating}.
Depending on the input precision (INT8 \textit{vs.} FP16), the psum output from the PEs is routed to the appropriate hardware resource within \flextree.
In \figureautorefname~\ref{flextree}, for $IC_P$ values of [1, 2], the flops [A, B, C, D, E, F, G, H] at level 1 serve as the final OF output tap points. For $IC_P$ = [4], the flops [I, J, K, L] at level 2 act as the final OF output tap points. Similarly, for $IC_P$ values of [8] and [16], the flops [M, N] at level 3 and [O] at level 4, respectively, serve as the final OF output tap points. Therefore, the total number of \flextree output tap points varies for different $IC_P$ values. Therefore, for $IC_P$ values of [1, 2, 4, 8, 16], the total number of \flextree output tap points is [8, 8, 4, 2, 1], respectively. To simplify the extraction of final OF points from the \flextree module into the drain module, we allow a maximum of four OF points to be extracted from \flextree in one round. The figure illustration assumes $IC=64, IC_P=16$, and therefore Port O is active.

As is evident from the above discussion, \flextree achieves dynamic reconfiguration of the depth of the adder tree. This configurable feature is aided by software-programmable configuration registers. Unlike existing DNN accelerators where partial sum accumulation occurs by moving psums among neighboring PEs, \flextree's innovative tree-based architecture significantly enhances partial sum accumulation efficiency (up to $2.14\times$ speedup). In contrast to state-of-the-art DNN accelerators with fixed schedules and adder tree-based architectures, where the adder tree depth remains fixed at design time, our \flextree technique offers dynamic reconfiguration capabilities, achieving speedups of up to $4\times$\nobreakdash--$16\times$, across seven DNNs, namely, ResNet50, GoogleNet, InceptionV2, MobileNetV2, MobileNetV3, SqueezeNet1.1 and MobileNet\_SSD. Thus, our proposed \flextree architecture enhances compute efficiency by allowing superior psum accumulation techniques across a wide range of layers found in modern DNNs.



\subsection{Schedule-aware Tensor Distribution Network (SDN)}
\label{sec:sdn}

The Schedule-aware Tensor Distribution Network (SDN) serves as one of the fundamental architectural innovations of the proposed \dnn accelerator, tasked with efficiently transferring input data between on-chip memory (SRAM) and flexible PE array, and vice versa, adhering to the optimal layer schedule \cite{chinya2024schedule}. These data include configuration settings, activation and kernel data (IF \& FL), sparsity encodings, as well as bias \& scale factors essential for calculation within the PE array. Additionally, the SDN manages the transportation of computational results, including output activation (OF) and partial sums, from the PE array's internal storage structure (RF) back to the SRAM, ensuring that the layout facilitates subsequent tensor layer acceleration. During the operational phases, the input side of the distribution network is termed the ``load/fill" phase, while the output side is termed the ``drain" phase. In fixed hardware accelerators, the pre-determined data layout in SRAM simplifies the load and drain phases, but compromises flexibility and optimization in operations. This rigidity restricts reuse potential, escalates memory accesses, and significantly increases overall energy and power consumption. Flexible hardware demands dynamic changes in the SRAM data layout, contingent on the type of reuse and optimal schedule (blocking and partitioning) for the layer.

\subsubsection{Load Path}
\label{sec:load}

The usual design consideration revolves around simplifying one of \textit{load} or \textit{drain} phase, while the other phase manages the complexities associated with rearranging the data to adhere to the optimal schedule. When the SRAM data layout remains fixed, the loading process must handle the complexity associated with unpacking the fixed layout data and arranging them according to the predetermined order and sequence dictated by the optimal schedule. Furthermore, the loading process must be hierarchical: initially organizing the input in a manner consumable by a column of PEs based on the partitioning factor and then, within the column, determining which input byte corresponds to which PE based on the blocking factor through a series of multiplexers \cite{mathaikutty2022sparsity, mathaikutty2022data}. For activation data, this involves retrieving data from memory in a predetermined order and distributing the IX, IY, and IC in the sequence and quantity specified by the reuse factor of the optimal schedule. Throughout the reuse process, one set of input remains resident, while the other circulates multiple times. Typically, the optimal schedule strives to maximize reuse, thereby reducing the frequency of fetching from SRAM. Ideally, a fully flexible DNN accelerator would allow partitioning in both the incoming activation and weights. However, this approach introduces a significant rise in MUXing complexity, along with its accompanying overheads, resulting in a convoluted routing process in the load, circular buffer, and PE FSM. To mitigate these challenges, we restrict weight partitioning within column and activation partitioning across column of the PE array.  This strategy aims to streamline routing complexities and enhance operational efficiency within the accelerator architecture.

\begin{figure}[t]
	\centering
	\includegraphics[width=\columnwidth]{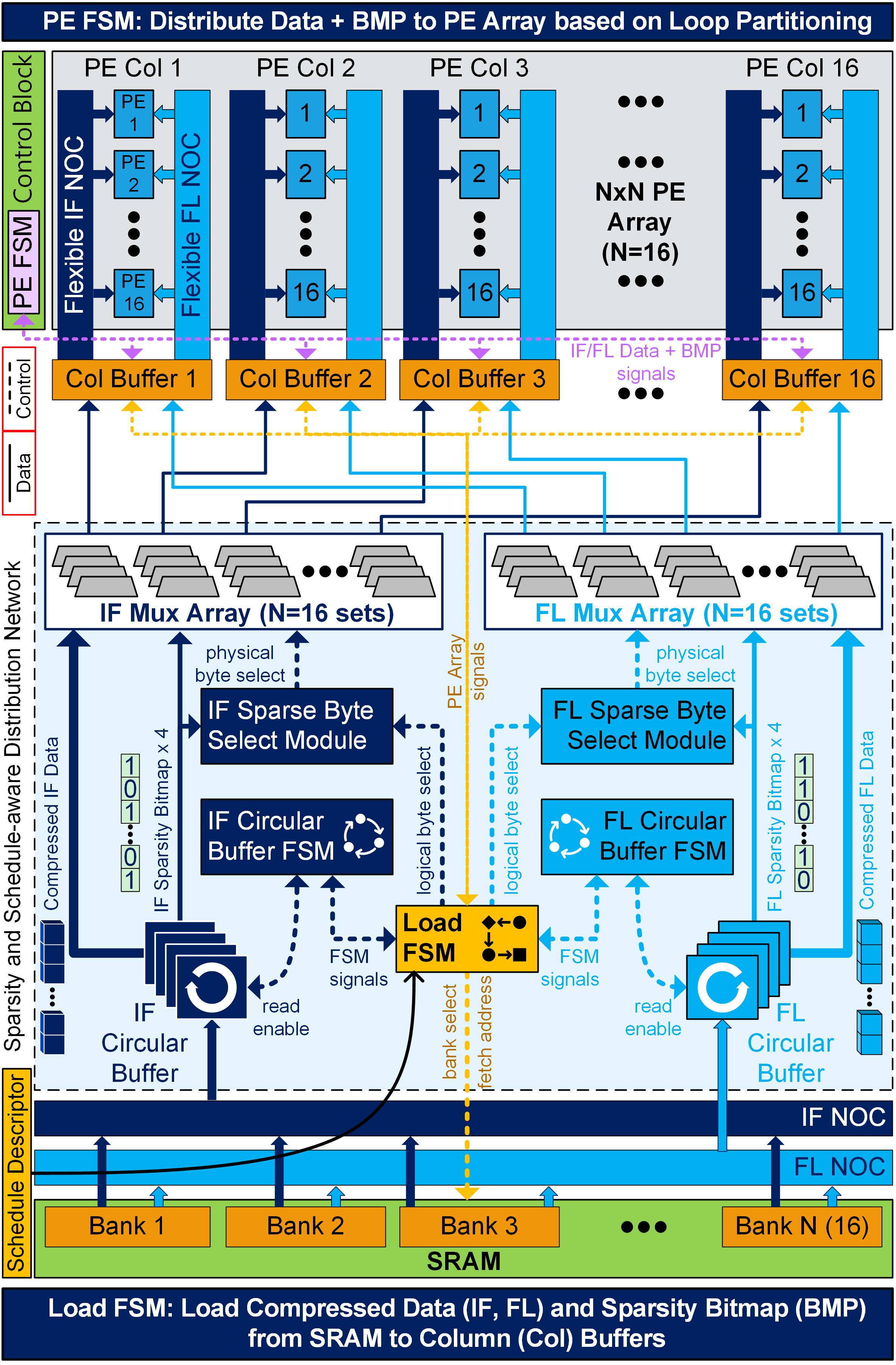}
	\caption{\dnn accelerator load path. Compressed activations (IF) and weights (FL) along with sparsity bitmaps are fetched from SRAM and delivered to PE array via the NoC interfaces. Flexible schedule support is integrated in Load FSM, Circular Buffer FSM, and PE FSM. OF NoC (part of drain) not shown inside PE array for clarity.}
	\label{load}
\end{figure}

\figureautorefname\ref{load} illustrates the crucial \textit{load} path of \dnn, spanning from SRAM to PEs. Central to this pathway is the \textbf{load FSM}, which interfaces with SRAM, the circular buffer FSM, sparse byte select modules, and PE columns. Activation of the load FSM is initiated by a $start$ signal, indicating that all configuration register values have been appropriately set by the schedule descriptor based on the optimal schedule. Optionally, this $start$ signal can also serve as the reset input for the load FSM, ensuring the removal of any outdated register values from previous layers.
Once the FSM is active and space becomes available within the circular buffer, the FSM transmits \textit{fetch address} and \textit{bank select} signals to SRAM, and the IF and FL NoCs commence the transmission of IF and FL data to their respective circular buffers. Concurrently, each PE within the PE columns returns credits to the load FSM as space becomes available in the any of the double-buffered IF/FL data RF. Upon receipt, the FSM directs read requests to the circular buffer FSM. Within the circular buffer FSM, metadata is managed and data are dequeued in response to load FSM requests, signaling when the buffer is empty.

In cases where valid compressed IF/FL data and corresponding sparsity bitmaps are present in the buffers, the circular buffer FSM initiates bitmap transmission to the sparse byte select module and compressed IF/FL data transmission to the IF/FL multiplexer (mux) array. The load FSM calculates \textit{logical byte select} signals based on current load counter values and configuration registers. These signals are utilized by the byte select modules, along with sparsity bitmaps, to determine \textit{physical byte select} signals, which control the IF and FL mux arrays for data routing, accounting for potential compression.
These mux arrays facilitate data routing between the circular buffers and the column buffers associated with each PE column. It is important to note that when communicating with the PE, control signals pass through the distribution network, utilizing a specific number of staging buffers to meet timing requirements. The Load FSM remains active, fetching data until the entire load volume is processed, at which point clock gating is initiated.

\begin{figure}[t]
	\centering
	\includegraphics[width=\columnwidth]{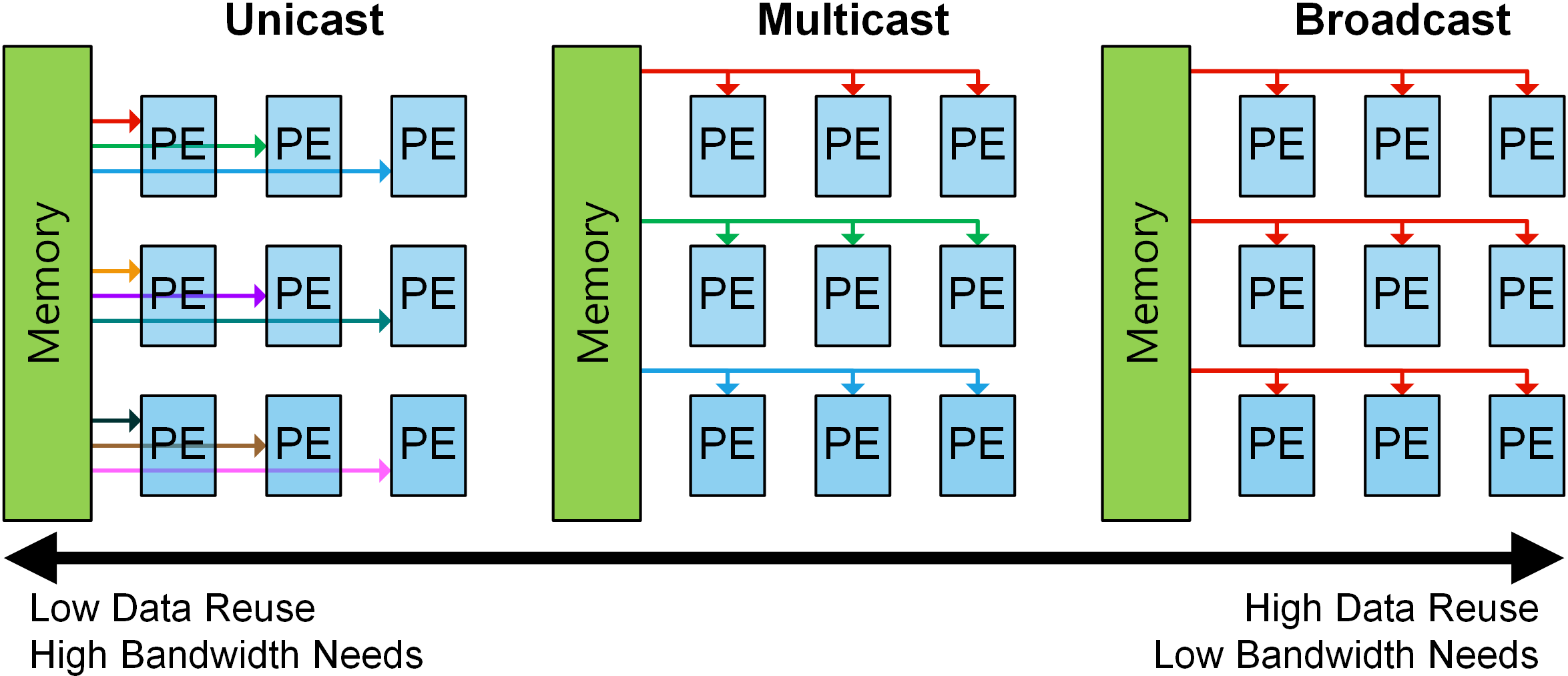}
	\caption{Data distribution patterns through flexible NoC.}
	\label{NoC}
\end{figure}

Another critical aspect of the distribution network is the interconnect or Network-on-Chip (NoC) used to link the PE array with the drain and load blocks in the design. NoC must have the ability to unicast, multicast, or broadcast the input data to one or more PEs based on the order specified by the optimal schedule, as demonstrated in \figureautorefname~\ref{NoC}. This maximizes reuse and minimizes the number of accesses to SRAM, improving overall efficiency, as defined in existing research \cite{kwon2018maeri}. As shown in \figureautorefname\ref{load}, when valid compressed data are loaded into column buffers, IF and FL NoCs distribute data to the appropriate PEs.

\subsubsection{Drain Path}
\label{sec:drain}

One of the core innovations within the \dnn architecture is \flexdrain, an efficient framework for processing OF maps across various schedules.  systematically drains OF maps, specifically tailored for flexible schedule-based DNN accelerators. Focusing on MAC operations along the IC dimension, the fixed drain pattern ensures consistent extraction of OF points in an IC-major fashion, regardless of the current layer schedule. This design choice capitalizes on the understanding that sparse compression in DNN accelerators predominantly occurs along the IC dimension. Implementing this fixed draining methodology simplifies drain design, integrating schedule awareness into load logic with minimal overhead. This novel approach holds promise for advancing DNN accelerators, enhancing reuse, reducing memory traffic, and improving energy efficiency.

\begin{figure}[t]
	\centering
	\includegraphics[width=\columnwidth]{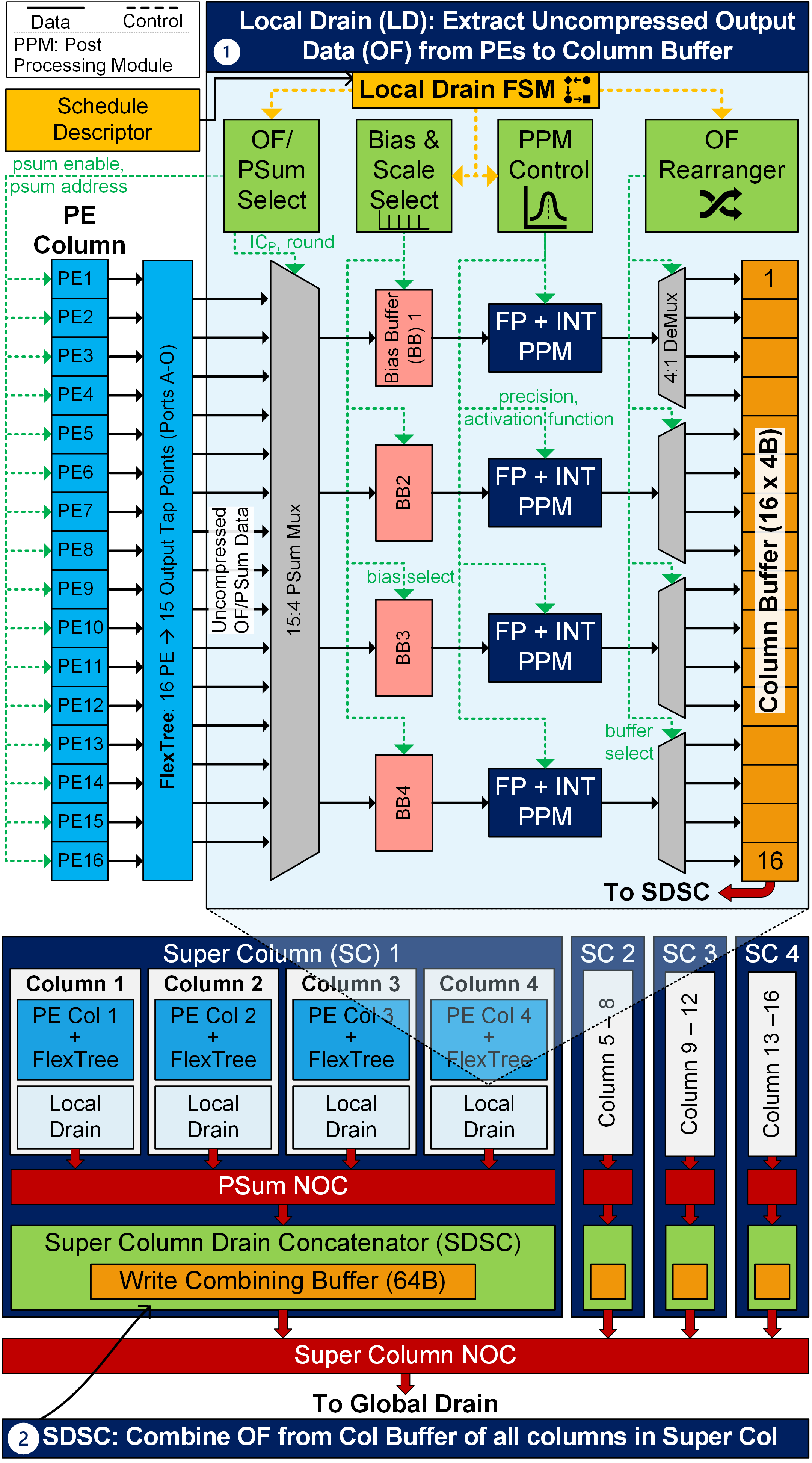}
	\caption{\dnn accelerator local drain path, showing (1) local drain path routing output activations/partial sums from each PE inside a PE column to column buffer. Group of 4 columns organized into super column. (2) SDSC per super column routes data from local drain to global drain.}
	\label{local_drain}
\end{figure}

The \flexdrain datapath encompasses several agents distributed across the PE compute array subsystem. \figureautorefname \ref{local_drain} provides a high-level depiction of these components and their respective functions. The components constituting the drain data path are: 1) Local Drain (LD): Instantiated on a per-column basis, the LD is responsible for extracting the output activation or psums from the PE, facilitating their transfer to the Super Column Drain Concatenator (SCDC). 2) SCDC: Implemented on a per-super column basis, the SCDC is tasked with concatenating data from the output column buffers of all 4 columns within a super column using psum-NoC. Subsequently, it transmits this concatenated data to the Global Drain (GD) via the super-column NoC. 3) GD: Serving as a central agent, it plays a pivotal role in rearranging SCDC data in a 1$\times$1$\times$Z manner, which further encodes/compresses these data and writes them to the SRAM. 

\noindent
\textbf{Local Drain:} The Local Drain (LD) operates to extract output activation data from the PEs within a column, forwarding them to the Post Processing Modules (PPMs), and then directing the PPM outputs to the column's output buffer, subsequently routed to the centralized GD. An overview of the local drain datapath is depicted in \figureautorefname \ref{local_drain}. The accompanying block diagram illustrates the various components of the Local Drain at a high level, each of which will be elaborated upon in subsequent sections. As previously outlined, each PE can generate up to 16 OF points per round for a given set of input data, with the exact count contingent upon the layer and input tensor parameters. Upon readiness, these OF points transition from the active OF RF to the shadow OF RF. Following this transfer, the PE signals to the Accumulate Finite State Machine (AccumFSM), a part of local drain FSM, that the associated Local Drain is primed to extract the OF points from the PEs.

\textit{Flow of Control:} Based on the layers and tensor parameters configured in the registers, it is possible to determine whether AccumFSM needs to consume data for accumulation across PEs, particularly if ICs are distributed across different PEs. In such scenarios, the LD waits for the AccumFSM to complete processing these OF points before proceeding. When AccumFSM is active, the LD streamlines the flow, refraining from extracting OF points until the accumulation is complete. In contrast, when AccumFSM is not required prior to LD operations, the PEs trigger the extraction process to the LD. The extraction sites and the number of points in each PE with a valid OF point to be extracted are determined using configuration registers. This information guides the sequential extraction of OF points from the shadow OF RF. Upon completing the extraction from the shadow OF RF, the LD indicates to the PEs that the shadow OF RF is fully utilized and prepared for the next round of transfers from the active OF RF.

\textit{OF Select to PPM:} The OF outputs accumulated from the \flextree flexible adder architecture (\sectionautorefname\ref{sec:flextree}), multiplexed (MUXed) using a 15:4 MUX, are fed into each PPM. LD orchestrates the selection of inputs for the psum-MUX in a round-robin manner, facilitating the transfer of input data into the PPM. The functionality of the PPM, primarily used for activation functions, quantization, \textit{etc.,} is configured via its bank of configuration registers, which serve as the foundation for processing the input data and selecting biases/scales. The LD assumes the responsibility of steering the data path for the PPM, issuing input data alongside bias/scale values, and subsequently extracting the output data to feed into the output column buffer. Note that the PPM module can handle both integer and floating-point precision.

\textit{OF Rearranger:} The PPM data output links to the column OF buffer entries through a 4:1 DeMUX, configured by LD FSM based on the drained OF point context. LD directs the PPM output to the appropriate buffer entry. In layers where certain PEs yield no OF points, LD ensures that 0 values populate the corresponding buffer entries, facilitating seamless data drainage by GD. For floating-point cases, data must be outputted in high-low pattern for seamless processing by both GD and Sparse Encoder.

Taking into account the area and performance specifications of the accelerator, it has been established that each column, comprising 16 PEs, should integrate 4 PPMs. This configuration includes 4 INT PPMs and 4 FP PPMs, activated when FPMACs are enabled. Each of these PPMs is exclusively allocated to serve 4 PEs, ensuring optimized resource utilization and efficient processing capabilities.

\noindent
\textbf{Super Column Drain Concatenator (SCDC):} The Super Column Drain Concatenator (SCDC) plays a pivotal role in consolidating data from the output column buffers of all 4 columns within a super column and forwarding it to the centralized GD via the Super Column NoC (SC-NoC). Each output column buffer within a column is 4 bytes wide and 16 entries deep. Once all round-required entries are filled in individual column buffers, LD in each column transmits 16 bytes of data from its column buffer to the SCDC through the psum agent packet NoC (psum-NoC) per round. SDSC combines these 4$\times$16B values into 64B data. A 2-bit super-column ID (SCID) is appended to create a 514-bit data packet for GD, which is subsequently transferred to the GD via the SC-NoC. Thus, the SDSC serves as a crucial link between the LD and the GD. Note that when PPM is enabled, each generated output activation is 1 byte for \textit{INT8} precision or 2 bytes for FP16/BF16 precision. In both cases, data from the column buffer is dispatched to the SCDC in 16-byte chunks. 

\begin{figure}[t]
	\centering
	\includegraphics[width=\columnwidth]{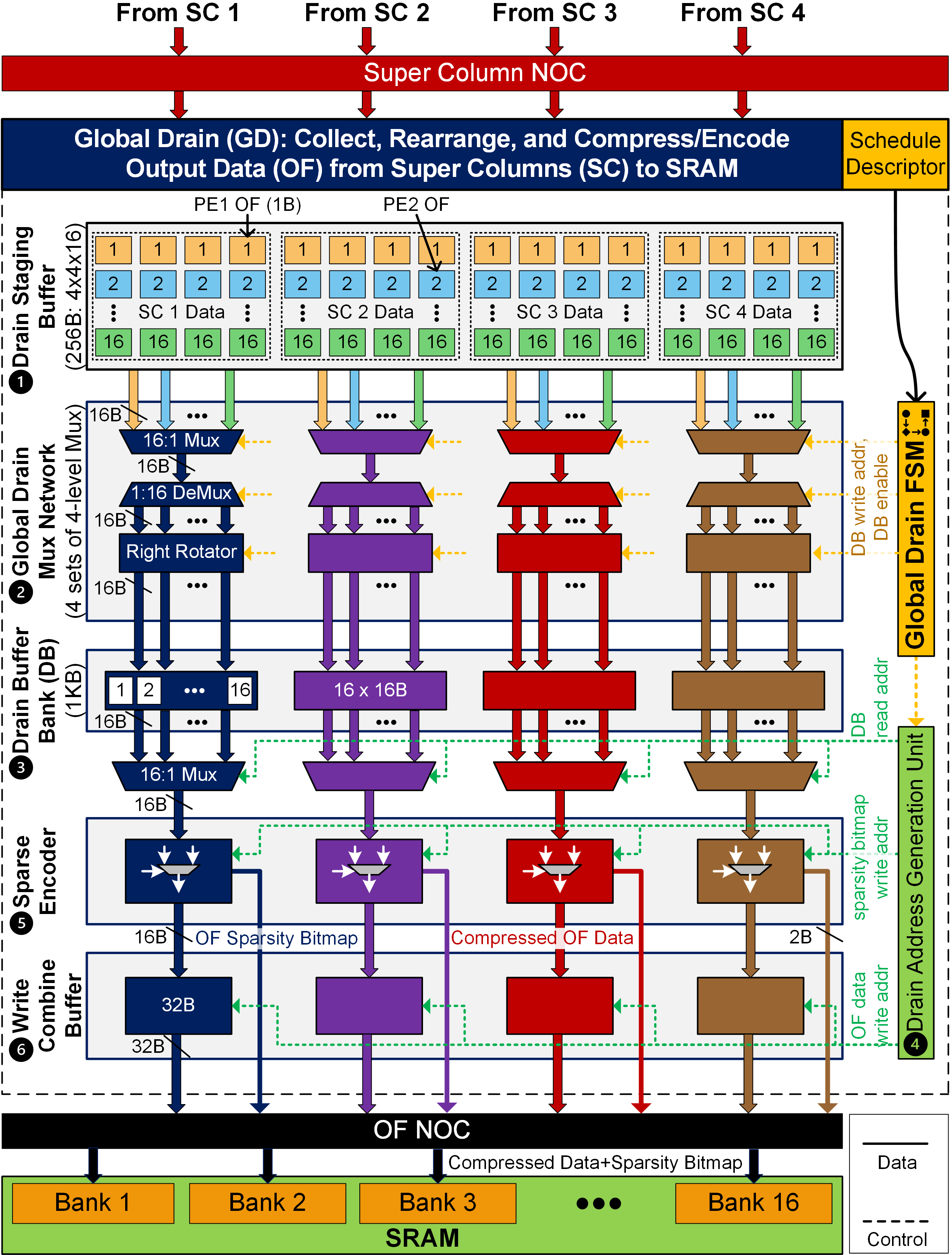}
	\caption{\dnn accelerator global drain path, showing (1) drain staging buffer, (2) global drain mux network, (3) drain buffer banks, (4) drain address generation unit, (5) sparse encoders, and (6) write combining buffers. Collectively, these units drain uncompressed data from local drains PE array and writes compressed data to SRAM.}
	\label{global_drain}
\end{figure}

\noindent
\textbf{Global Drain:} The Global Drain (GD), as demonstrated in \figureautorefname\ref{global_drain}, serves as the central hub within the PE array subsystem, tasked with gathering output activation points from PEs across all columns. Its primary function involves rearranging these activations into a 1$\times$1$\times$Z format, where Z represents the OC dimension for the current layer (or the IC dimension for the subsequent layer). Subsequently, the GD encodes these activations and writes them to the SRAM for further processing or storage. The GD comprises the following components: 
1) A 256B input buffer called the Drain Staging Buffer (DSB) where the output activations from all PEs are staged. 64B data from each Super Column are transmitted to the GD via the SC-NoC, and concatenated into the DSB, thereby generating 256B for processing. 
2) Global Drain Mux (GDM) network consisting of 4 sets of multiplexer arrays that rearrange the staged output activations from DSB into the Drain Banks. 
3) 64 Drain Banks (DB) organized as 4 groups of 16 entries, with each Drain Bank of size 16B, for a total of 1KB. These buffers serve as a pre-final staging area for output activations from PEs before encoding and writing to SRAM. Each 16B bank can hold 16 OCs for the next layer, controlled by GDM for writing and DAGU for reading.
4) Drain Address Generator Unit (DAGU), which computes the ${x,y,z}$ coordinates of the points in the DBs that will be written into the SRAM. The read signals drive a drain bank reader (not shown in figure) to get the data out of the DBs. 5) A bank of four Sparse Encoders (SE), which encode the data to be written to the SRAM. 6) 4 Write combining buffers that allow the writing of compressed data and the corresponding sparsity bitmaps into the SRAMs.
Among the six components, the Global Drain Mux and Sparse Encoder are the most important elements of the GD. Detailed explanations of these components are provided below.

\textit{Global Drain Mux}: The role of the GD Mux control logic is to manage the selection process across the various mux stages to drain data from the DSB. There are a total of four GDMs, each capable of independently accessing DSB entries but restricted to writing solely to its designated group of DBs. Let us dive into each stage of the GDM: \textit{Stage S1: Entry Select:} This stage involves choosing 16B of the DSB data. Following the configuration registers, the GD Mux control logic adopts a row-wise selection approach from the DSB, utilizing a 16:1 entry selects Mux capable of picking any of the 16 row entries. \textit{Stage S2: Bank Select:} The 64 drain banks (DBs) are grouped into four sets, each containing 16 entries. The bank select function determines to which of the 16 entries the data will be written to. Notably, data can be written to multiple entries, potentially all 16 entries in an extreme scenario. The bank select or bank enable ensures proper multicasting of data to the DBs. \textit{Stage S3: Right Rotator:} After each GDM multicasts the selected DSB entry to one or more DBs, it assigns an appropriate right rotation value specific to each DB. This step is crucial to \textit{align and concatenate consecutive Zs within a single DB}. \textit{Stage S4: Byte Enable:} Finally, byte enable serves as the write byte enable, ensuring that the correct set of bytes from the selected DSB line is written to the DB.

\begin{figure}[t]
	\centering
	\includegraphics[width=0.87\columnwidth]{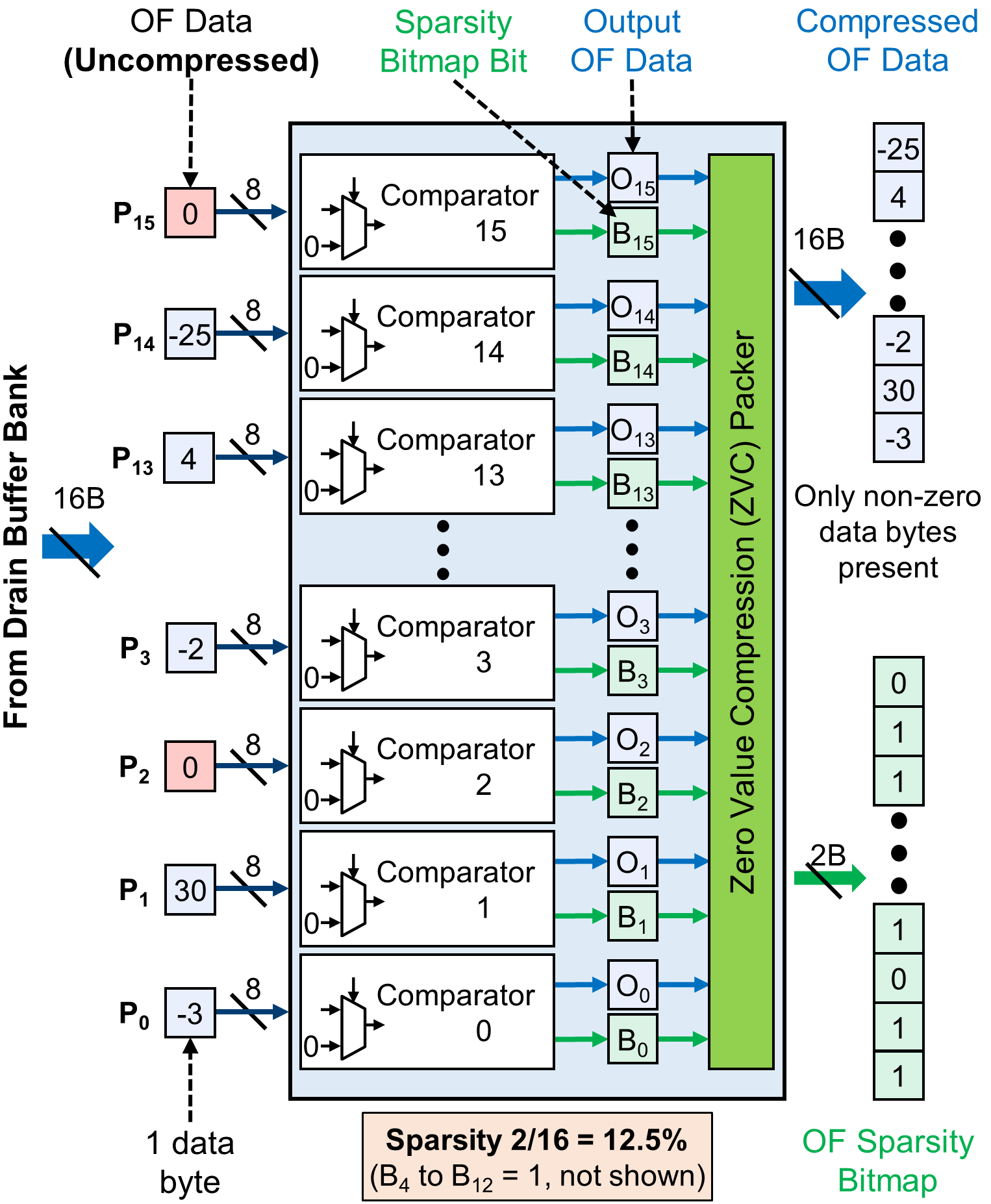}
	\caption{Sparse encoder \cite{ghosh2024harvest} in global drain performing zero-valued compression with illustration.}
	\label{sparse_encoder}
\end{figure}

\textit{Sparse Encoder}: This is a pivotal element within the global drain, crucial for leveraging data sparsity to enhance the speed of inference processing in \dnn. \figureautorefname\ref{sparse_encoder} provides a schematic representation of the SE block. Its primary function is to compress dense data streams by discarding zero values, thereby outputting a compressed data representation accompanied by a sparsity bitmap. The drain buffer acts as a staging area for the data before they are written to SRAM, providing the SE with input. Each DB bank is allocated to store data corresponding to a unique output context (OX, OY) point, with a 16B payload containing the OC data for that specific OX, OY coordinate pair. A single context, representing one data stream, may extend across multiple banks and up to 16 contexts can be processed concurrently within a group.
The GDM ensures that banks flagged with valid bits, indicating that their data has yet to be processed by the SE, are protected from being overwritten. The SE itself operates on 16B granularity, compressing the data for each distinct context contained within the DB. The degree of sparsity dictates the number of input lines that the SE must handle before it can produce a compressed 16B output line for a particular context. Alongside the compressed data, the SE generates a unique sparsity bitmap for each context, which is then sent to SRAM through the OF-NoC. The address for writing the bitmap to SRAM is determined by DAGU. As elements within a context stream may be received over several cycles, the SE is designed to manage context switching efficiently by preserving the state of each context and retrieving it when necessary to continue processing.

\subsection{Two-sided Sparsity Acceleration}
\label{sec:csal}

In the proposed \dnn accelerator, our aim is to harness the sparsity in both FLs and IFs to enhance not only \textbf{energy efficiency} but also \textbf{DNN inference throughput}. Throughout the accelerator, the data remains compressed until reaching the PE. Operating within the compressed domain offers advantages, such as \textit{reducing on-chip bandwidth requirements and storage demands}. However, handling compressed data, which often varies in length, poses challenges in terms of data manipulation, such as distributing data across PEs and implementing sliding window processing within the PE \cite{raha2022system}. In this section, we will introduce an innovative \textbf{two-sided sparsity acceleration logic} capable of processing sparse data within the compressed domain to achieve higher throughput \cite{chinya2021accelerated, raha2021methods, raha2022methods, kundu2024rash}. This logic spans multiple units including VPE, load path, and drain path.

The core idea is that IFs or FLs with a value of $0$ do not contribute to non-zero outcomes during MAC operations, allowing them to be skipped during both the compute and storage phases \cite{connor2020dot}. As explained in \sectionautorefname\ref{sec:load}, SRAM serves as storage for zero-value compressed input activations (IF) and weights (FL), which are delivered to each column buffer in batches through the load path in SDN \cite{chinya2023methods, raha2021methods}. The PE FSM then transmits the compressed IF and FL to their respective buffers (CD RF) in each PE. Along with these, the corresponding bitmaps are also transferred to IF and FL sparsity bitmap buffers, respectively. 

\begin{figure}[t!]
    \centering
    \includegraphics[width=1\columnwidth]{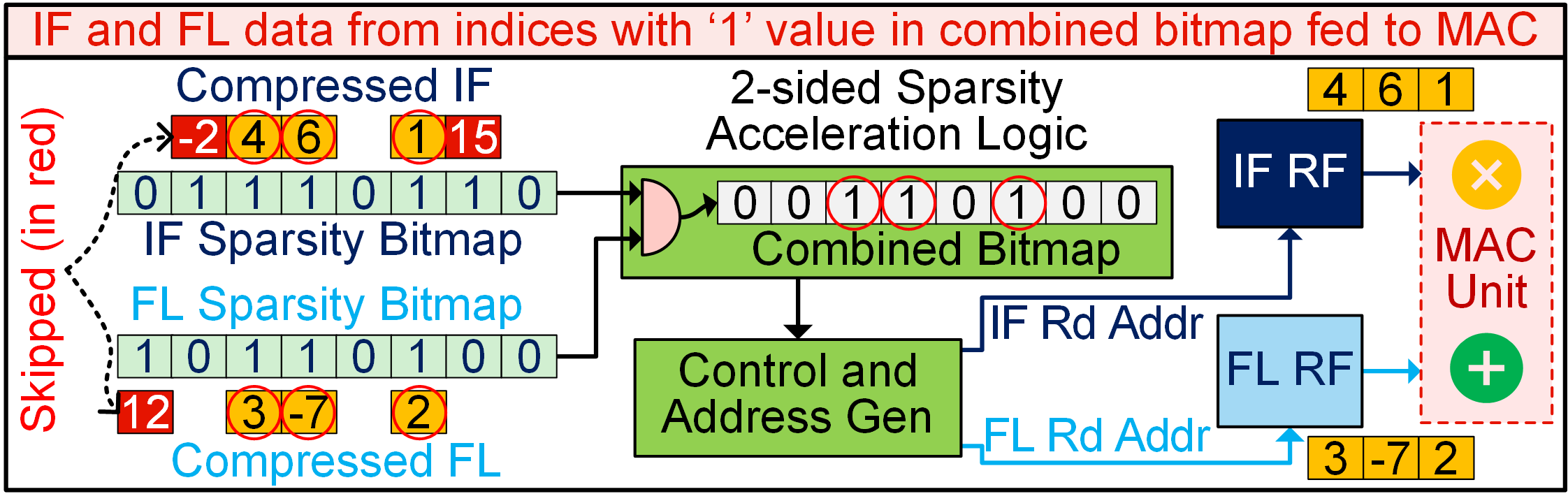}
    \caption{Two-sided combined sparsity acceleration logic.}
    \label{fig:csal}
\end{figure}

As illustrated in \figureautorefname\ref{flexarray}, the two-sided sparsity acceleration module receives the bitmaps as input. The bitmaps consist of 1-bit values, represented by either `0' or `1', instead of 8-bit values present in the original IF and FL sets (considering an 8-bit quantized network). For every non-zero element in IF, the corresponding position in the activation bitmap consists of `1'. The bitmap consists of a `0' for every zero value in the incoming IF set. The FL sparsity bitmap is also generated in an identical way. Subsequently, through a series of combinational operations, it determines the combined sparsity bitmap indicating exact non-zero positions in IF and FL. The total number of $1$s in this intermediate bitmap depicts the total number of activation and weight pairs that need to be computed in the MAC unit and result in non-zero partial sums, which must be accumulated over time. Through the CAG unit, these values are then fed from the IF and FL RF into the MAC unit, which generates the OF maps, as demonstrated in \figureautorefname~\ref{fig:csal}. Subsequently, these feature maps flow through the drain path, which incorporates a zero-value compression module (in SE) to compress the zero-valued elements. These compressed feature maps are then stored back in the SRAM (or DRAM) for further processing in subsequent layers of the DNN. Through the combined sparsity acceleration logic, \dnn achieves enhanced computational speed, improving performance, and throughput for DNN inference by exploiting two-sided sparsity. This approach significantly decreases the energy consumption of the PE array, completing tasks with a reduced number of cycles.

%% file: 4_exp_method.tex
\section{Experimental Setup}
\label{sec:exp_method}

\begin{figure}[t!]
	\centering
	\includegraphics[width=0.9\columnwidth]{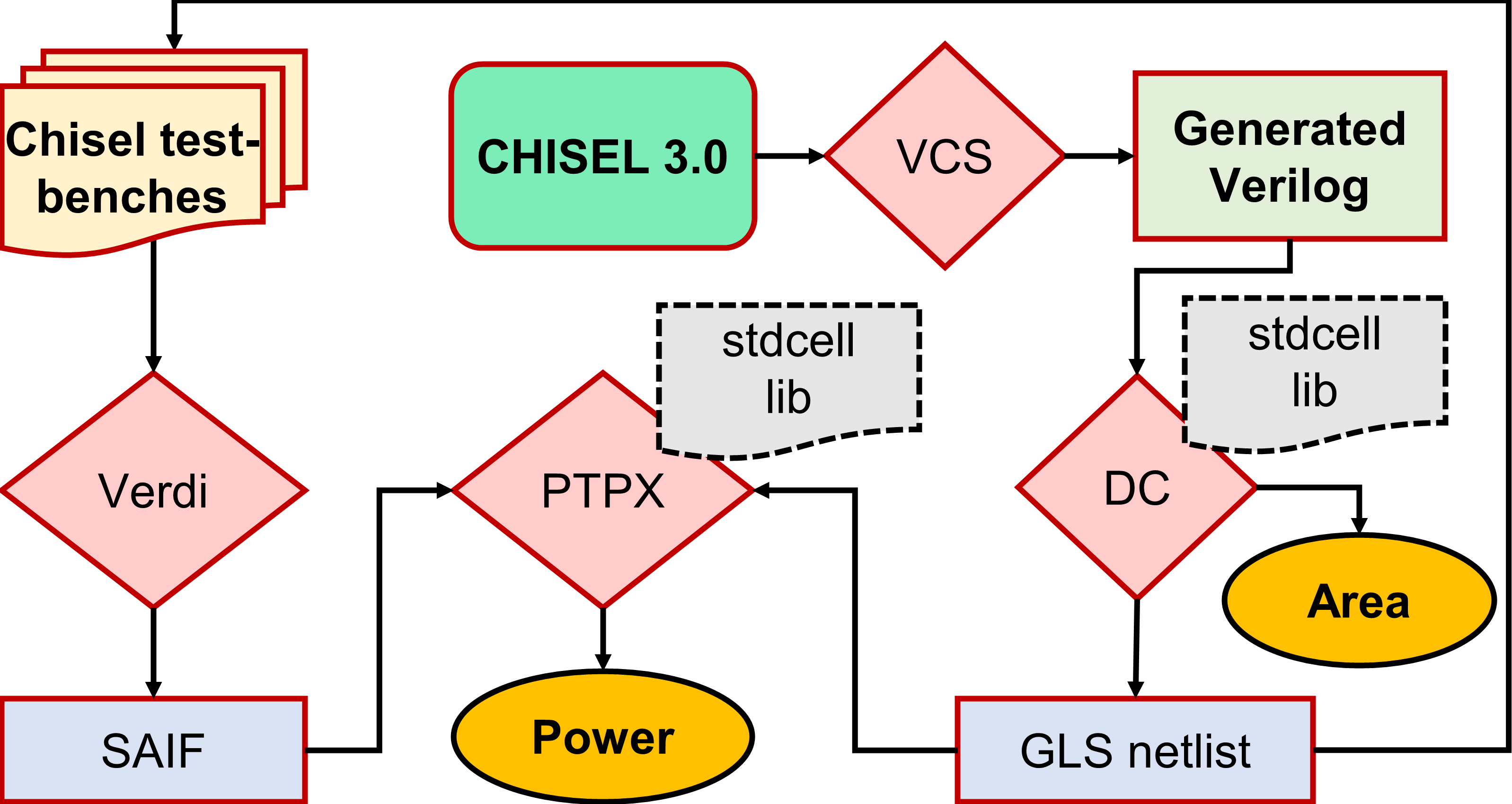}
	\caption{Workflow overview for \dnn implementation. Please refer to text for full-form of abbreviations.}
	\label{chisel}
\end{figure}

\begin{table*}[tp]
    \centering
    \caption{Comparison of \dnn with state-of-the-art fixed schedule accelerator designs. Identical memory hierarchies and cost ratios are used for evaluation.}
    \begin{tabularx}{1.95\columnwidth}{>{\RaggedRight}m{0.6\columnwidth}XXX} 
        \toprule
        & \textbf{Eyeriss \cite{eyeriss}} & \textbf{TPU \cite{tpu}} & \textbf{\dnn} \\
        \midrule
        \textbf{Memory Hierarchy} & 3-level\parnote{Eyeriss has additional inter-PE communication with RF:PE=1:2 cost ratio} & 3-level & 3-level \\
        \textbf{Num of PEs} & 168 & 256 & 256 \\
        \textbf{RF (in each PE)} & 512 B & 32 B & 208 B \\
        \textbf{On-chip Buffer/SRAM\parnote{SRAM sub-bank size remains constant for all}} & 108 KB & 64 KB & 1.5 MB \\
        \textbf{DRAM} & 1 GB & 28 MB & 1 GB \\
        \textbf{Energy cost ratio} \textbf{\mbox{(PE:RF:SRAM:DRAM)}} & 1:1:6:200 & 1:0.06:6:200 & 1:0.125:6:200 \\
        \bottomrule
    \end{tabularx}
    \parnotes
    \vspace{-15pt}
    \label{tab:accel}
\end{table*}

To evaluate the efficiency of the proposed accelerator, we implemented \dnn in Chisel 3.0, with the generated RTL simulated in Synopsys VCS®. We chose Chisel because of its ability to facilitate the generation of parametrizable designs featuring multiple variations of PE, allowing easy modification of RF size, number of MAC units, etc.
As mentioned in \sectionautorefname~\ref{sec:dnnacc}, the accelerator supports \textit{UINT8, INT8, FP16, BF16} precision.  
Subsequently, the RTL undergoes synthesis in the Synopsys Design Compiler (DC), utilizing one of the industry’s most advanced process technology nodes, (based on 7 nm), to generate the Gate-Level Netlist (GLS) and corresponding area for each accelerator component.
To estimate the power consumption within the proposed \dnn accelerator, we employed Synopsys Verdi to generate an activity file (Switching Activity Interchange Format: SAIF), using test benches for assistance. The accelerator netlist, coupled with the activity file, serves as input to Synopsys PrimeTimePX (PTPX), enabling power estimation at the gate level for both block and full-chip designs of the \dnn accelerator. An overview of this workflow is illustrated in \figureautorefname~\ref{chisel}.
The \dnn architecture comprises a unified tile of 256 PEs organized in a 16×16 grid (16 columns with each column having 16 individual PEs), featuring 8 MAC units within each PE, resulting in a total of 2048 MACs. This tile encompasses 1.5 MB of SRAM equipped with 32-byte read/write ports. The PE consists of 4x16 B IF Data RF Register File (RF), 4x16 B FL Data RF, and 16x4 B OF RF. In addition, each PE also consists 4x2B IF sparsity bitmap RF and 4x2B FL sparsity bitmap RF, which is 1/8th the size of data RF as 1 bit in bitmap is used to represent 1 byte in data. Together, these RFs contribute to 208B RF per PE. The precision of the IF, FL, and OF points is an 8bit integer. The memory hierarchy of our design is illustrated in Table \ref{tab:accel}. Operating at a frequency of 1.8 GHz and 0.75 volts, the accelerator boasts a dense peak Trillion Operations Per Second (TOPS) performance, reaching 7.37 TOPS, with efficiency metrics of 5 TOPS/watt and 4.6 TOPS/mm$^2$.

We conducted a comparative analysis of the performance of \dnn in conjunction with two state-of-the-art dense accelerators, namely Eyeriss \cite{chen2016eyeriss} and TPU \cite{jouppi2017datacenter}. The comparison considered various design specifications described in Table~\ref{tab:accel}. 
Furthermore, we evaluated the performance of \dnn on sparse DNN workloads using state-of-the-art networks \cite{paperWithCodeWebsite}: \resnet, \mobilenet, \inception, and \gnet trained on the ImageNet dataset \cite{krizhevsky2012imagenet}. The first three models were compressed using (i) Quantization-Aware Training (QAT) to quantize weights/activations to INT8 precision and (ii) unstructured pruning using the regularization-based sparsity algorithm (RB-sparsity). \gnet was quantized in the same way, but filter pruning with geometric median criterion was applied. The compressed models were obtained from Intel's Neural Network Compression Framework (NNCF)\cite{kozlov2020neural}. Per-layer and overall network weight sparsity were obtained from these models. Furthermore, all models were subjected to inference on the entire ImageNet2012 validation dataset (50,000 images) and activation sparsity at input and output of each layer was calculated using PyTorch hooks. The average activation sparsity across the entire dataset, weight sparsity, and layer statistics were fed into a framework of \dnn, which was used to obtain the layer-wise and overall network compute acceleration and total energy consumption of the accelerator, reported in \sectionautorefname~\ref{sec:results}. Specifically, we compared the performance of \dnn, which uses two-sided combined sparsity, against dense accelerators without any sparsity support and those capable of exploiting fixed weight-sided sparsity \cite{chen2016eyeriss, park2020optimus}. The framework was modified to evaluate the latency and energy of a dense variant and a weight-sided variant of \dnn to allow fair comparison.

%% file: 5_results.tex
\section{Experimental Results}
\label{sec:results} 

In this section, we begin by providing a breakdown of the power and area consumption for our proposed \dnn accelerator. Subsequently, we proceed to evaluate its performance using state-of-the-art network and dataset configurations.

\subsection{\dnn Power and Area Results}

\begin{figure}[t!]
	\centering
	\includegraphics[width=0.95\columnwidth]{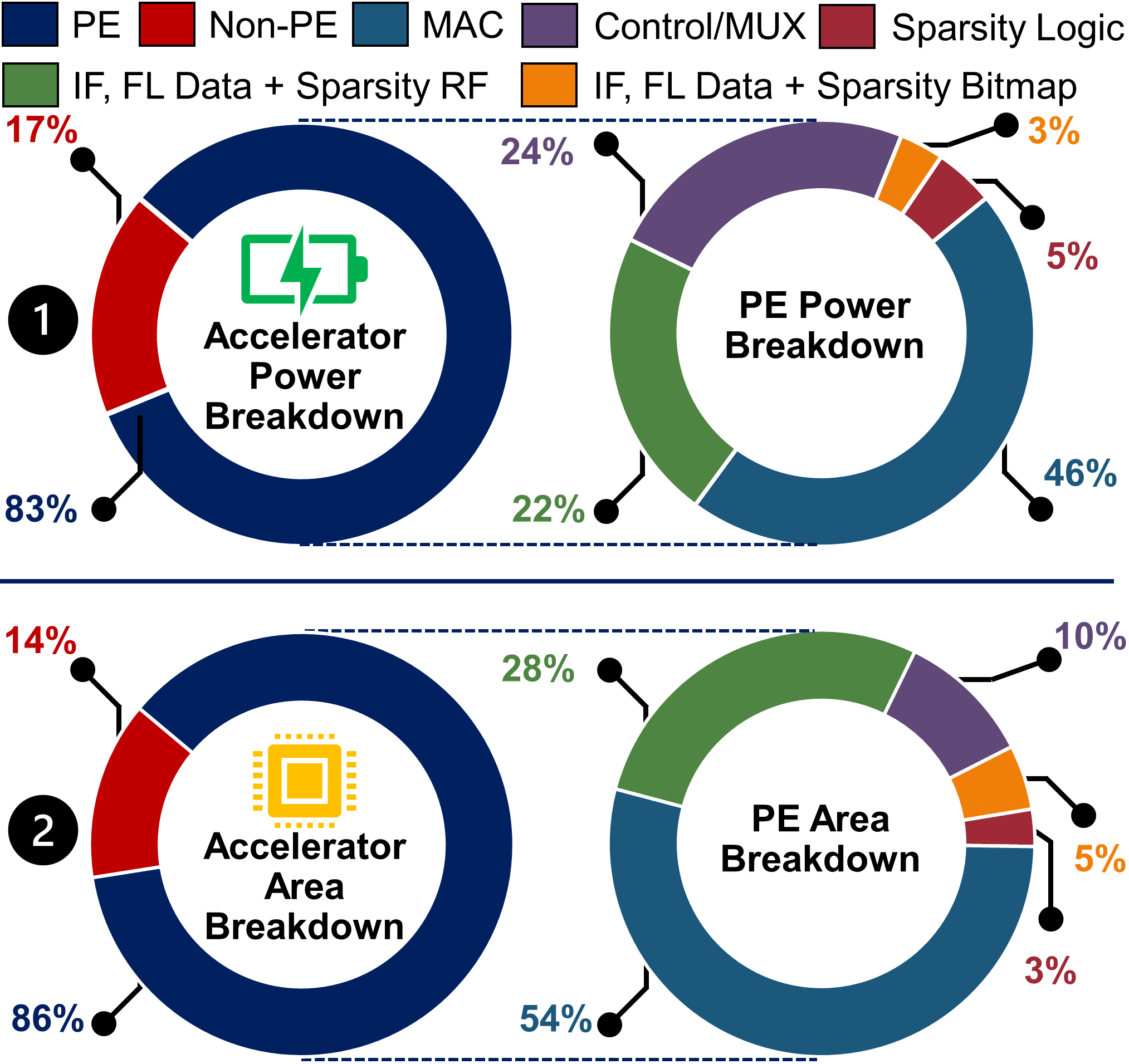}
	\caption{(1) Power and (2) Area Breakdown of \dnn for overall accelerator and PE level granularity.}
	\label{area_power}
\end{figure}

We assess the power and area cost of the proposed \dnn accelerator using an illustrative implementation in this paper. \figureautorefname~\ref{area_power}.1 and 2 show the power and area breakdown of the entire accelerator as well as the inter-PE breakdowns, respectively. As shown, the PE array unit consumes $\approx 83\%$ of power and $\approx 86\%$ of the total area of the overall accelerator design. Furthermore, the MAC operation constitutes about $46\%$ of the power and $54\%$ of the area inside each PE of the \dnn accelerator. This shows reasonable power and area impact compared to the significant benefits it provides in terms of the ability to support flexible schedules.


\subsection{Comparison with SOTA Fixed Schedule Accelerators}

\begin{figure}[t]
	\centering
	\includegraphics[width=\columnwidth]{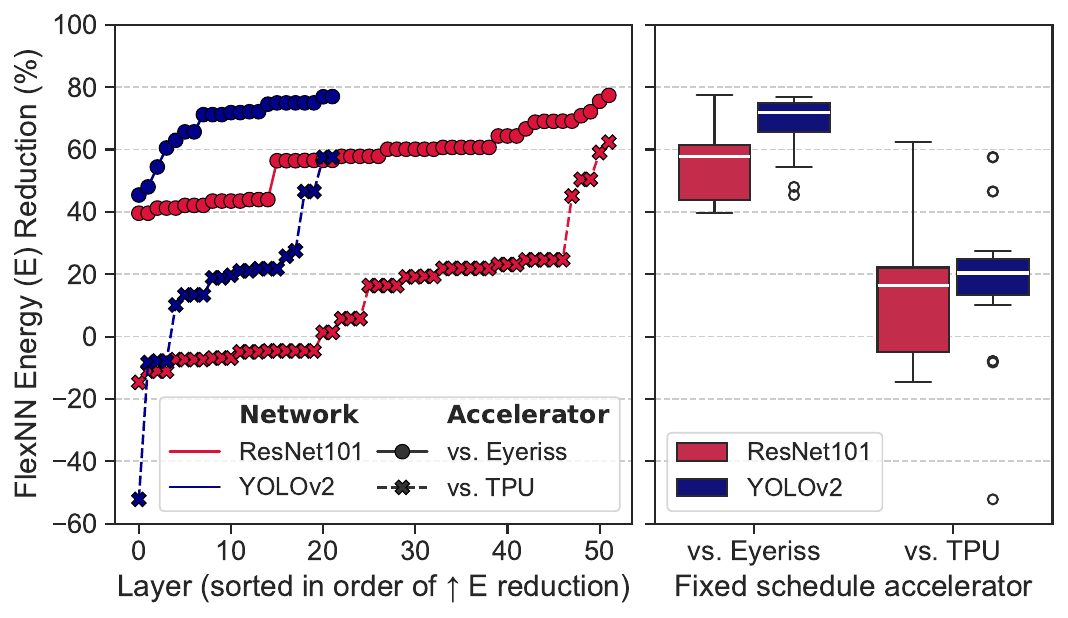}
	\caption{Layer-wise distribution of \% energy improvement of \dnn over fixed schedule DNN accelerators (Table \ref{tab:accel}) for ResNet101 and YOLOv2 (dense models). Optimal schedule used for each layer in \dnn accelerator.}
	\label{energy_reduction}
\end{figure}

\figureautorefname~\ref{energy_reduction} shows the improvement in energy efficiency of our flexible schedule DNN accelerator \dnn over two prominent fixed-schedule designs, Eyeriss~\cite{eyeriss} and TPU~\cite{tpu} assuming identical memory hierarchies. These results are obtained for two DNNs used in image classification and object detection, ResNet101 and YOLOv2, using our custom DNN accelerator energy estimation framework. We have used dense models (i.e., with 0 weight sparsity) for these results. Note that we have scaled the memory hierarchy of the two accelerators to the same level as \dnn for a fair comparison. In this figure, Here, the y-axis represents a \% reduction in energy consumption of \dnn compared to these two designs. The left subplot depicts the layer-wise energy reduction for all layers, sorted in increasing order of reduction. In the right subplot, we summarize the distribution of reduction across all layers. The x-axis shows the two comparative accelerators. Compared to Eyeriss, \dnn results in $40\%$\nobreakdash--$77\%$ reduction for ResNet101 and $45\%$\nobreakdash--$77\%$ for YOLOv2. Compared to TPU, \dnn provides up to $62\%$ and $58\%$ energy savings for ResNet101 and YOLOv2, respectively. 
While it is true that in certain layers, \dnn exhibits a slight energy increase (indicated by a negative energy reduction) compared to TPU, this is primarily attributed to the optimized dataflow in TPU for specific layers, particularly 20 layers in ResNet101 and 4 layers in YOLOv2. However, on average, \dnn still provides notable advantages, offering average energy savings of $14\%$ and $22\%$ for these respective DNN architectures over TPU. It is important to note that the increased energy consumption in these layers comes from the robust support for flexibility within DNN, which inherently introduces slightly higher overhead. On the other hand, we see an average improvement of $57\%$ and $69\%$ over Eyeriss. Despite occasional spikes in energy consumption for select layers, \dnn consistently outperforms these fixed-schedule accelerators, showcasing its superior efficiency and overall cost-effectiveness.

\subsection{Sparsity Benefits using \dnn}

In this section, we present a comprehensive analysis of the layer-wise and overall network speed-up achieved by \dnn compared to two prominent counterparts: a dense accelerator without any sparsity acceleration support and a fixed weight-sided sparse accelerator. \figureautorefname~\ref{layerwise_speedup}.1-4 presents the layer-wise compute acceleration (y-axis) provided by weight-sided and \dnn in comparison with the dense accelerator, for few representative layers (x-axis) of 4 DNN benchmarks. Note that the activation sparsity numbers reported in the following discussion are averaged across the entire dataset. For a fair comparison, benchmarking was performed using the same optimal schedule for all accelerator types.

\subsubsection{\resnet}
The sparse \resnet model has $5\%$\nobreakdash--$88\%$ unstructured weight sparsity, $weight\_sp_{layer}$, resulting in up to $8.1\%$ acceleration across layers in weight-sided accelerator. However, except before the first conv layer, \resnet has a high activation sparsity, $act\_sp_{layer}$,  at the input of every convolution layer due to the presence of the ReLU activation function. On average across the entire ImageNet validation dataset, this amounts to $act\_sp_{layer}$ = $14\%$\nobreakdash--$83\%$ sparsity. \dnn conveniently leverages both weight and activation sparsity to provide up to $10.3\%$ compute acceleration, as shown in \figureautorefname~\ref{layerwise_speedup}.1. \textbf{Overall, \dnn gives up to $3.1\times$ better acceleration than the weight-sided accelerator for \resnet}. 

\subsubsection{\gnet}
Since \gnet was filter-pruned, maximum $weight\_sp_{layer}$ = $30\%$. This contributed to maximum $1.4\times$ speed-up in weight-sided accelerator. In contrast, the maximum measured $act\_sp_{layer}$ = $91\%$ resulted in a maximum acceleration $10.8\times$ in \dnn. \textbf{\figureautorefname~\ref{layerwise_speedup}.3 shows that \dnn provides up to $7.7\times$ better compute acceleration compared to fixed weight-sided accelerator, even for networks with low weight sparsity.}

\subsubsection{\inception}
This model is very sparse with a maximum $weight\_sp_{layer}$ = $96\%$. There are many layers with large dimensions and filter sizes; therefore, both the weight-sided accelerator and \dnn can leverage weight sparsity and provide up to $24.7\times$ speed-up for \textit{Mixed.7a.branch3x3.2.conv, Layer Id: 72} (not shown in figure). Although $act\_sp_{layer}$ for this layer is $78\%$, \dnn cannot provide any additional speed-up for this layer. However, there are many other layers with activation sparsity higher than weight sparsity, allowing \dnn to leverage both. As depicted in \figureautorefname~\ref{layerwise_speedup}.4, \dnn provides a high level of compute acceleration. Among such layers with sparsity skewed toward activations, the maximum speed-up is $11.3\times$. Therefore, the proposed design can take the best of both worlds and give better savings than the weight-sided accelerator. \textbf{Across all layers, \dnn is up to $4.3\times$ faster than the weight-sided accelerator, clearly demonstrating superior performance.}

\subsubsection{\mobilenet}
\begin{figure*}[th!]
	\centering
	\includegraphics[width=2\columnwidth]{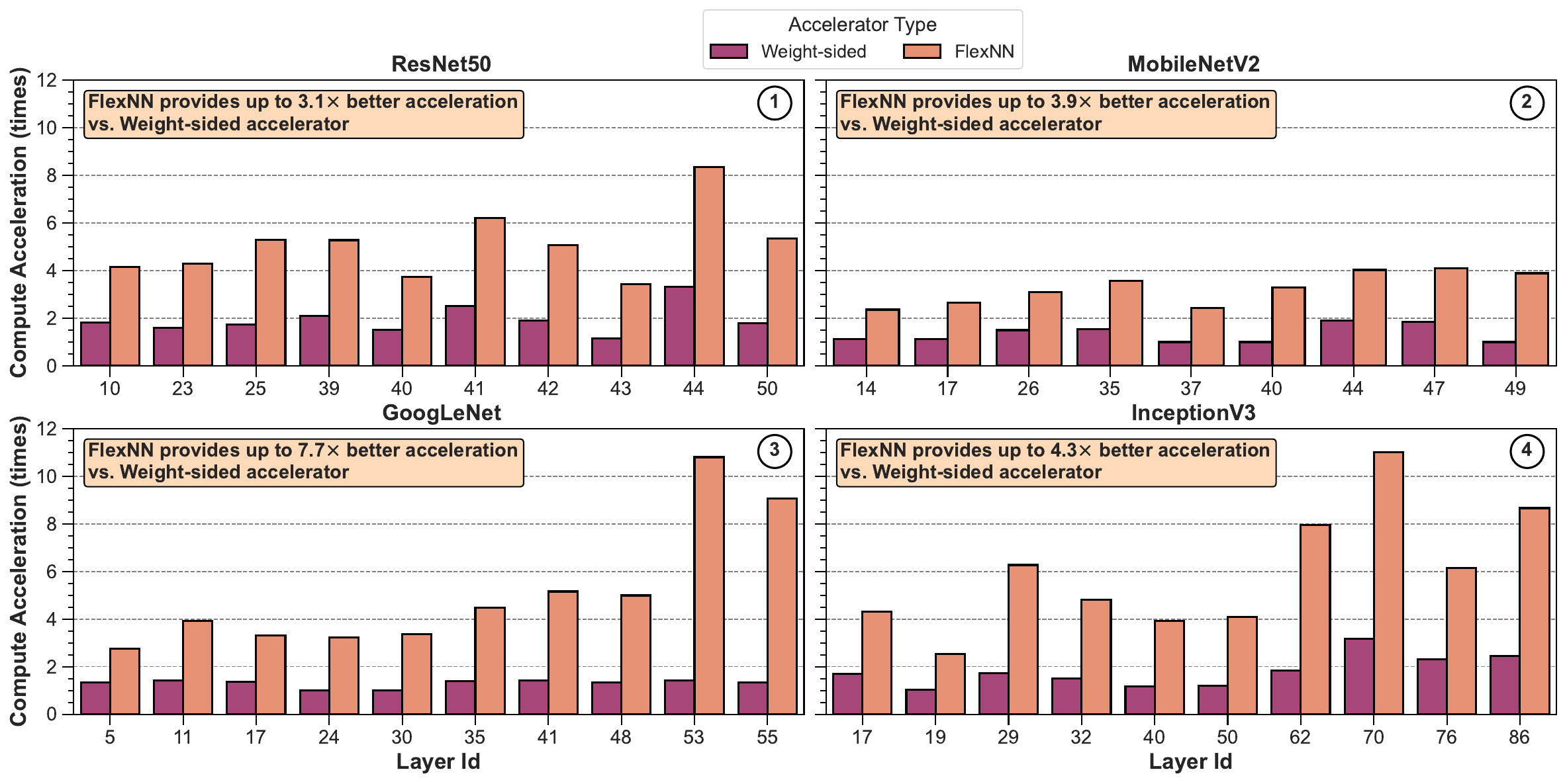}
	\caption{Comparison of layerwise compute acceleration of \dnn and Weight-sided (one-sided) sparse accelerator over dense accelerator (without any sparsity support) benchmarked with (1) \resnet, (2) \mobilenet, (3) \gnet, (4) \inception. Only a few representative layers are presented for each network.}
        \vspace{-10pt}
	\label{layerwise_speedup}
\end{figure*}

\mobilenet is a compact and lightweight model compared to the other benchmarks discussed earlier. Although sparse \mobilenet consists up to $70\%$ $weight\_sp_{layer}$, the maximum speed-up provided by the weight-sided accelerator is only $3.3\times$ (last linear layer). Interestingly, the weight sparsity of all conv layers, except \textit{features.18.0, Layer Id: 51} is $<50\%$ leading to a low overall speed-up. However, \dnn leveraging activation sparsity (maximum $act\_sp_{layer}$ = $74\%$) in addition to weights can provide up to $4.1\times$ acceleration. \textbf{\figureautorefname~\ref{layerwise_speedup}.2 indicates that even for compact models with small layer sizes, \dnn is superior to the weight-sided accelerator by $3.9\times$.}

\begin{figure}[t]
	\centering
	\includegraphics[width=0.95\columnwidth]{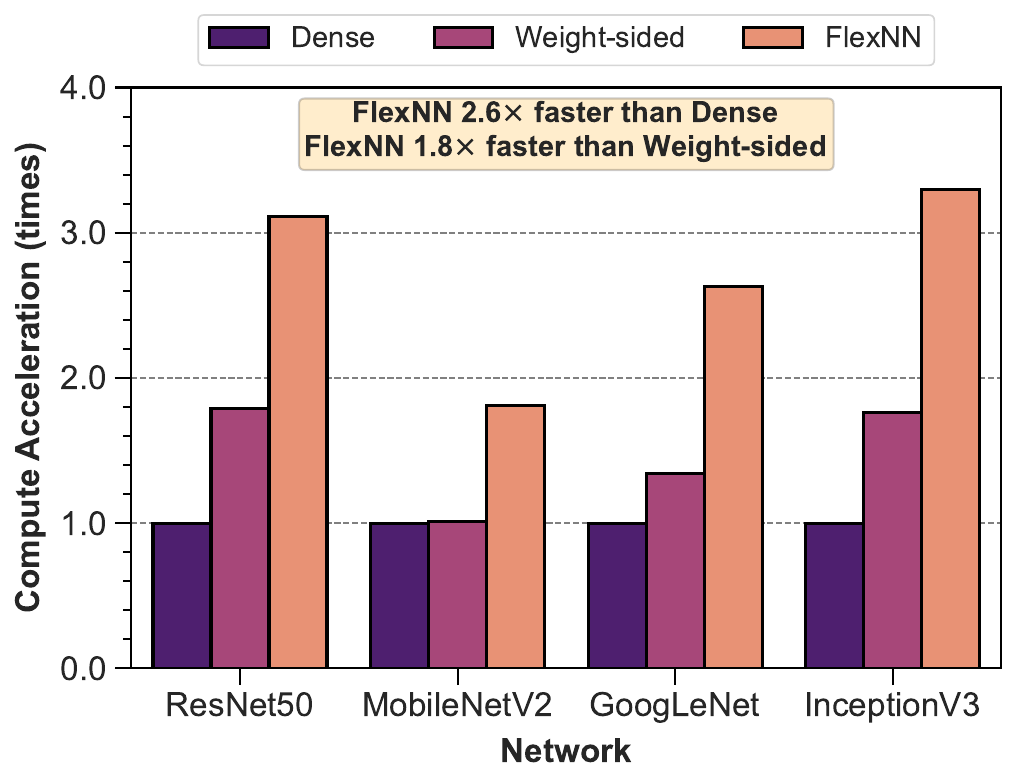}
	\caption{Comparison of compute acceleration for full network inference in \dnn over dense and weight-sided accelerator benchmarked with 4 DNNs.}
	\label{flexnn_speedup1}
\end{figure}

\subsubsection{Overall Network Acceleration}

The compute acceleration obtained by dense, weight-sided and our proposed accelerator for the entire end-to-end network inference, depicted in \figureautorefname~\ref{flexnn_speedup1} reveal a significant acceleration advantage conferred by \dnn across all evaluated networks. Here, the y-axis represents the acceleration, whereas the x-axis represents benchmarks. The dense accelerator does not provide any acceleration as it cannot leverage weight or activation sparsity, denoted by values $1$. Evidently, the speed-up for weight-sided accelerator is proportional to the overall network weight sparsity. 
Across all these networks, the weight-sided accelerator provides $1.01\times$\nobreakdash--$1.79\times$ speed-up. On the contrary, the acceleration obtained in \dnn is proportional to the relative distribution of weight and activation sparsity. 
For \resnet, $weight\_sp_{network}, act\_sp_{network}$ = $61\%, 55\%$ and \dnn takes advantage of them to provide $3.11\times$ speed-up. 
\mobilenet and \gnet has $weight\_sp_{network}, act\_sp_{network}$ = $52\%, 30\%$ and $weight\_sp_{network}, act\_sp_{network}$ = $24\%, 58\%$, respectively. These results in $1.81\times$ and $2.63\times$ speed-up in \dnn, respectively. Clearly, even with these two networks with low sparsity on one side, \dnn provides a significant amount of computation due to two-sided sparsity support. 
Finally, \inception has $weight\_sp_{network}, act\_sp_{network}$ = $61\%, 63\%$ contributing to $3.3\times$, which is the maximum across the 4 networks. 
As evident from these results, our accelerator consistently outperforms both dense and weight-sided architectures in terms of compute acceleration.\textbf{ This substantial improvement, $2.6\times$ \textit{vs.} dense and $1.8\times$ \textit{vs.} weight-sided accelerator (\textit{geomean}), underscores the efficacy of our proposed approach in enhancing overall network speed-up, demonstrating its superiority in accelerating DNN inference computations. }Furthermore, the observed acceleration benefits are valid across the various architectural complexities and model sizes represented by the diverse DNNs considered in our evaluation. This robust performance underscores the versatility and effectiveness of two-sided sparsity acceleration support in \dnn across a spectrum of DL model architectures.

\subsubsection{Energy Efficiency improvement}

\begin{figure}[t]
	\centering
	\includegraphics[width=0.95\columnwidth]{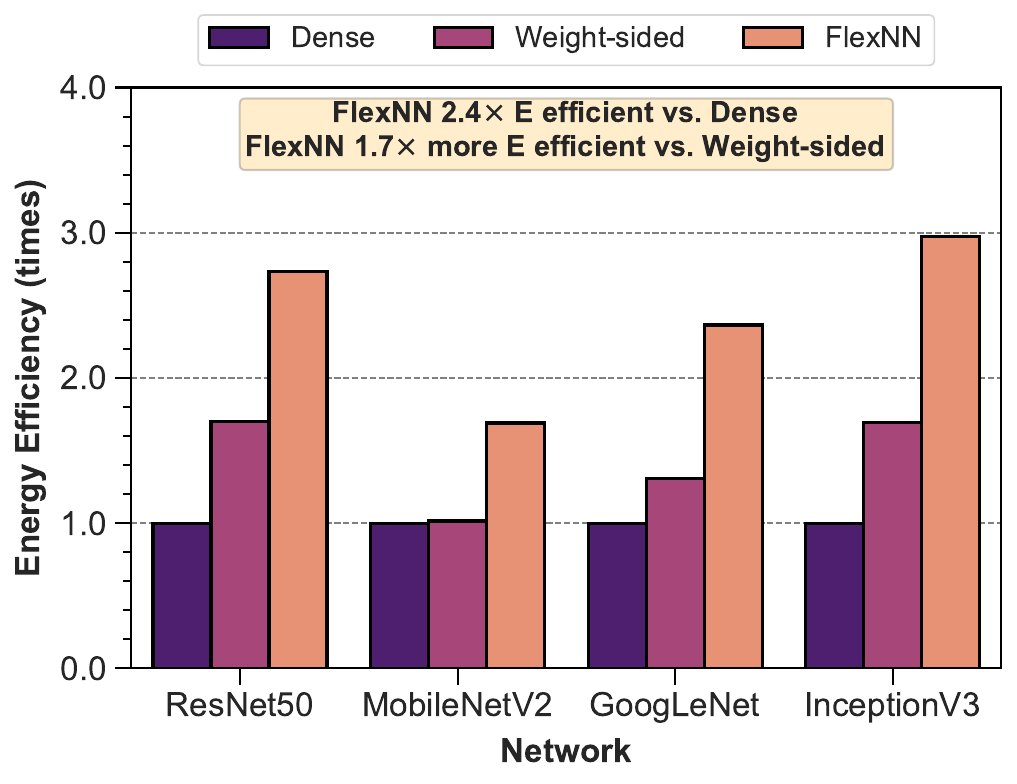}
	\caption{Comparison of energy efficiency for full network inference in \dnn over dense and weight-sided accelerator benchmarked with 4 DNNs.}
	\label{flexnn_energy_savings}
\end{figure}

\figurename~\ref{flexnn_energy_savings} presents the improvement in energy efficiency of 3 different accelerator architectures (y-axis) while evaluating 4 DNN benchmarks (x-axis) on the ImageNet validation dataset. We considered dense accelerator energy consumption as the baseline. As evident from the figure, these results largely correlate with overall network compute acceleration in \figureautorefname~\ref{flexnn_speedup1} since the accelerator circuits are active for a reduced amount of time. Furthermore, compared to the weight-sided accelerator, \dnn allows for substantial reduction in memory cycle count as ZVC compressed data flows through the different memory hierarchies, resulting in reduced memory energy consumption. This is enabled by the sparsity-aware load and drain path, as explained in \sectionautorefname~\ref{sec:sdn}. Note that DRAM transactions are not considered in these results. \textbf{Across all 4 benchmarks, \dnn is $2.4\times$ and $1.7\times$ more energy efficient than the dense and weight-sided accelerators, respectively.}

In conclusion, our comprehensive evaluation showcases not only the substantial compute acceleration achieved by our proposed accelerator, but also its remarkable energy efficiency improvements compared to existing dense and weight-sided architectures. This underscores the pivotal role of our approach in addressing the dual challenges of performance enhancement and energy conservation in DNN accelerators, paving the way for sustainable and efficient AI hardware solutions.

%% file: 6_related_work.tex
\section{Related Work}
\label{sec:related_work}

\begin{figure}[t]
	\centering
	\includegraphics[width=\columnwidth]{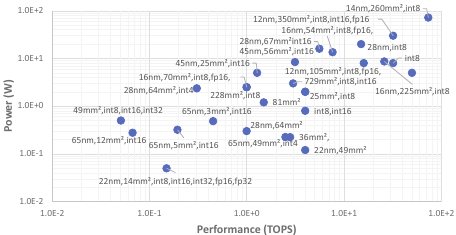}
	\caption{Edge DNN accelerator competitive landscape, plotted with public data, considering baseline reference~\cite{cost}.}
	\label{survey}
\end{figure}

Recent years have witnessed a surge in the field of DNN accelerators. Most DNN accelerator designs only implement fixed schedules with fixed dataflow. \figureautorefname~\ref{survey} illustrates the different DNN accelerators from industry and academia and their supported datatypes. For example, NeuFlow~\cite{neuflow} and ISAAC~\cite{isaac} implement a weight stationary schedule, ShiDianNao~\cite{shidiannao} and Movidius VPU2~\cite{kmb} implement an output stationary schedule, Google TPU~\cite{tpu} only implements Nonlocal Reuse schedule, and Eyeriss~\cite{eyeriss} from MIT implements a row stationary schedule. A key challenge arises from the limitations of the tensor data PE module hardware, which operates solely on a fixed dataflow pattern. It lacks the ability to dynamically adjust to accommodate diverse schedules, as it lacks awareness of any schedule information, owing to its restricted functionality.
Therefore, one cannot implement different schedules (i.e. dataflows) in these accelerators, and till today there are no existing accelerators that can support flexible schedules. In addition to hardware solutions, software-based solutions can mimic programmable PE array units that can perform computation on varying dataflow tensor data in general-purpose CPUs and GPUs, but fixed-function accelerators do not support this flexibility in design. Therefore, these software solutions cannot be used in existing accelerators. Moreover, software solutions are far from being energy optimal to be considered for adoption in edge inference devices.
FPGAs provide an alternative avenue for DNN acceleration with flexibility, but the hardware configuration of the FPGA cannot be changed during the execution of one DNN application, which also implies a fixed schedule during execution. Additionally, FPGAs have lower energy efficiency compared to ASIC hardware accelerators.

Since the PE array modules in all previous designs have limited functionalities in the form of a basic MAC structure, which is the compute kernel for convolution operation, the wide degree of mismatch between the data patterns required by different input, weight, and output stationary optimal schedules makes it impossible for the fixed architecture PE array to be able to handle the tensor data correctly without sacrificing energy and/or performance. The key disadvantage is that the PE array that performs convolution computation in these previous solutions is not schedule aware. Due to this limitation, reformatting the energy optimal dataflow type into a dataflow that is supported by the underlying fixed architecture PE array induces severe performance and energy penalty as more SRAM reads are required to complete the work and prevents the PE array from reaching maximum utilization if the accesses are serialized.  Software solutions can also be used for rearranging the input, output activation as well as weight tensor data for different optimal schedules to be fed into the PE array, into the type of fixed dataflow supported by the PE array, which not only would require assisting CPUs but would also be highly energy and performance inefficient, thereby significantly diminishing the energy efficiency gains offered by flexible scheduling.

In contrast to these approaches, we propose\textbf{ a schedule-aware runtime configurable PE array module}, which can (1) process tensor data (both weights and activations) that are either input, output, or weight stationary, or even a mixture of these dataflow types, depending upon the energy optimal schedule for the current DNN layer, (2) have its microarchitecture be reconfigurable at runtime based on software-programmable configuration registers, and (3) leverage maximum activation and weight reuse by having small amount of distributed local storage close to compute within the PE array itself. The dataflow support in the proposed PE array module is flexible. It is controlled by a list of configuration descriptors, which are set at the beginning of the execution of each layer. This tensor data computation PE array module is a pure hardware solution that exposes hardware knobs to the compiler and configures the dataflow during runtime, enabling the flexible schedules of convolutional layers in DNN accelerators without performance penalty due to rearranging computation within the PE or having to offload any work to CPU or software.

%% file: 7_conclusion.tex
\section{Conclusion}
\label{sec:conclusion}

In this paper, we proposed a flexible schedule-aware DNN accelerator \dnn, which can adapt its internal dataflow to the optimal schedule of each layer in DNNs. Our proposed solution maximizes data reuse at each memory level, resulting in significant energy savings arising from optimal data reuse. Note that flexibility works seamlessly on top of existing performance-enhancing features such as sparsity acceleration and low-precision logic, and it does not diminish their impact in any manner. It is evident that this flexibility comes at the cost of additional area overhead compared to fixed dataflow accelerators, but it also enables us to achieve significant energy savings on average across a myriad of DNN layers. 
Furthermore, we propose a novel approach to improve throughput and reduce energy usage in the \dnn architecture. Taking advantage of fine-grained sparsity in both activation and weight tensors, we optimize the inference engine within the hardware accelerator. Experimental results demonstrate significant improvements in both performance and energy efficiency compared to existing DNN accelerators. This research contributes to ongoing efforts to develop more efficient hardware accelerators for executing deep neural networks.

\balance